\documentclass[%
cp,  
 amsmath,amssymb,
 preprint,%
]{revtex4-1}

\usepackage{graphicx}
\usepackage{dcolumn}
\usepackage{bm}

\usepackage[utf8]{inputenc}
\usepackage[T1]{fontenc}
\usepackage{mathptmx} 

\usepackage{hyperref}
\usepackage[caption=false]{subfig}

\begin{document}

\title{Performance of GAN-based Augmentation for Deep Learning COVID-19 Image Classification}

\author{Oleksandr Fedoruk} 
 \email[Corresponding author: ]{oleksandr.fedoruk@ncbj.gov.pl}
\author{Konrad Klimaszewski}%
 \email{konrad.klimaszewski@ncbj.gov.pl}
\author{Aleksander Ogonowski}%
\author{Rafał Możdżonek}%
\affiliation{
  Department of Complex Systems, National Centre for Nuclear Research, Otwock-Świerk, Poland
}

\date{\today} 

\begin{abstract}
The biggest challenge in the application of deep learning to the medical domain is the availability of training data.
Data augmentation is a typical methodology used in machine learning when confronted with a limited data set.
In a classical approach image transformations i.e. rotations, cropping and brightness changes are used.

A promising avenue to solve this problem is the usage of Generative Adversarial Networks (GANs) to generate additional images 
that will increase the size of training datasets.
A GAN is a class of unsupervised learning where two networks (generator and discriminator) are joined by a feedback loop to compete with each other.
As a result, the generator constantly learns how to better deceive the discriminator, on the other hand, the discriminator gets constantly better at detecting synthetic images.

In this work, a StyleGAN2-ADA model of Generative Adversarial Networks is trained on the limited COVID-19 chest X-ray image set. 
After assessing the quality of generated images they are used to increase the training data set improving its balance between classes.

We consider the multi-class classification problem of chest X-ray images including the COVID-19 positive class that hasn't been yet thoroughly explored in the literature.
Results of transfer learning-based classification of COVID-19 chest X-ray images are presented.
The performance of several deep convolutional neural network models is compared.
The impact on the detection performance of classical image augmentations i.e. rotations, cropping, and brightness changes are studied.
Furthermore, classical image augmentation is compared with GAN-based augmentation.
The most accurate model is an EfficientNet-B0 with an accuracy of $90.2 \%$, trained on a dataset with a simple class balancing.
The GAN augmentation approach is found to be subpar to classical methods for the considered dataset.
We make our codebase publicly available\footnote{\href{https://github.com/cis-ncbj/covid19-stylegan-augmentation}{https://github.com/cis-ncbj/covid19-stylegan-augmentation}}.

\end{abstract}

\maketitle

\section{Introduction}

Simultaneously with the development of imaging techniques, progress in diagnosis support methods is made.
Currently, the leading approach for machine learning assisted classification of medical conditions is the utilisation of Convolutional Neural Networks (CNNs).
Examples of such usage are the detection of breast tumours in mammography imaging~\cite{Gardezi2019-ph,Rodriguez-Ruiz2019-jg} or detection of lung tumours in CT scans~\cite{Ardila2019-io,Kadir_2018,Jassim_2022}.
One challenge in the application of deep learning to the medical domain is the availability of training data.
For instance, some medical procedures are too expensive to perform and gather an appropriate set of data while some diseases are very rare.
An example of such a limited dataset is a collection of chest X-ray scans of patients who contracted COVID-19 disease caused by a strain of coronavirus called the Severe Acute Respiratory Syndrome Coronavirus 2 (SARS‐CoV‐2) that originated in Wuhan in the Hubei province in China. Especially at the beginning of the pandemic available data was limited, for example, the data set used by Hall et al.~\cite{Hall2020} consisted of 135 chest X-rays of COVID-19 and 320 chest X-rays of viral and bacterial pneumonia.
With enough time it is possible to aggregate a dataset large enough for deep learning applications.
Unfortunately, even then equal balancing of classes is not guaranteed which can lead to possible biases.
For example, NIH Chest X-rays dataset~\cite{Wang_2017} contains 112,120 X-ray images of 30,805 patients with fourteen common disease labels.
The largest category is Infiltration (25366 images) and the smallest is Hernia with only 284 images.
Another example is MRNet Dataset~\cite{Bien2018-dv} containing 1,370 knee MRI exams performed at Stanford University Medical Center. The dataset consists of 1,104 (80.6\%) abnormal exams, with 319 (23.3\%) ACL tears and 508 (37.1\%) meniscal tears.


Numerous works studied the applicability of X-ray chest scans in diagnosing COVID-19~\cite{Gouda_2022}.
Commonly the disease is detected using tests based on the reverse transcription‐polymerase chain reaction (RT‐PCR).
They are known to have high specificity but variable sensitivity for disease detection~\cite{Rubin2020-nd}.
On the other hand, while not recommended as a primary diagnostic tool, chest radiography and computed tomography (CT) scans are reported to be less specific than RT‐PCR but highly sensitive in detecting COVID‐19~\cite{Bai2020-pi}.
This leads to interest in the development of automated solutions that facilitate swift referrals for the COVID-19-affected patient population in a pandemic situation where resources are limited.

According to reports in the literature, COVID‐19 often manifests as ground‐glass opacities (GGO), with peripheral, bilateral, and predominant basal distribution in the lungs, causing breathing difficulties along with hyperthermia.
First to our knowledge attempt to design an AI-based detection method for COVID-19 from lung imaging were works by Xu et al.~\cite{Xu_2020} and by Gozes et al.~\cite{Gozes_2020}.
Both were geared to binary classification of COVID-19 and normal images like many works that followed~\cite{Sahinbas_2021,Medhi_2020,Alazab_2020,Narin2021-wp,Rajaraman_2020,Saad2022-uz,Islam_2021,Ezzat_2021,Sahlol_2020,Elshennawy_2020}.
The observed patterns in the lungs for COVID-19 disease are distinct yet visually similar to ones caused by viral pneumonia not related to COVID-19 or other pathogens~\cite{Rubin2020-nd}.
A study of literature shows difficulties in discerning viral pneumonia from cases caused by bacterial and fungal pathogens~\cite{Kermany2018-ok}.
Therefore a limited dataset leads to problems in the classification of images especially when one considers not only distinguishing between healthy and patients with COVID-19 but also patients with viral pneumonia and lung opacity~\cite{Khan_2020}.
In contrast to numerous works on the binary classification of COVID-19 X-ray images, there are several attempts to tackle the problem of distinguishing COVID-19 from other diseases affecting the respiratory system.
First, there is a group of works on the three class problem where viral pneumonia cases are included~\cite{Ucar_2020,Apostolopoulos2020-el,Houssein_2022,Gouda_2022}.
To our knowledge, the CoroNet~\cite{Khan_2020} was the first proposed model that includes four classes and obtains $89.6 \%$ accuracy.
A recent result by Khan E. et al.~\cite{KhanE_2022} achieved $96.13 \%$ accuracy utilising a modified EfficientNet-B1 model.

A classical approach to tackle problems with limited datasets is data augmentation~\cite{AsurveyonImageDataAugmentation}.
That is a technique to increase the data sample size by generating slightly modified copies of existing data or by generating completely new synthetic data.
In the case of image data, this amounts to various transformations like rotations, cropping, stretching or colour space transformations.
They are successful in increasing sample variation and safeguarding trained models from data collected in different conditions i.e. lighting or framing.
However resulting augmentations are still limited to a significant extent by the original dataset as most transformations will not affect the relationship between relevant features present in the image.
In particular, this limits their usefulness in scenarios with imbalanced data samples.

\begin{figure}
    \centering
    \includegraphics[width=0.7\textwidth]{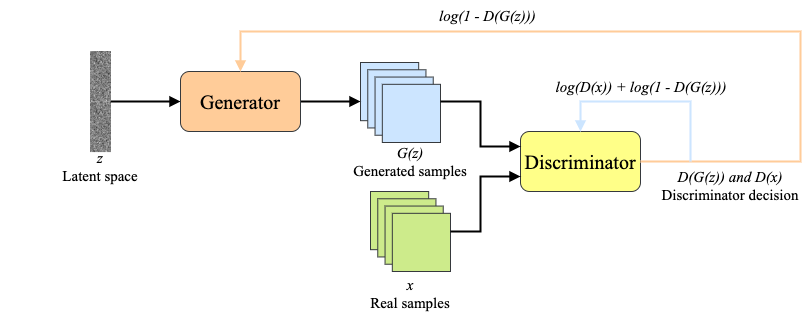}
    \caption{\label{fig:gan_architecture}A typical architecture of a Generative Adversarial Network.}
\end{figure}

A promising avenue to solve this problem is the usage of Generative Adversarial Networks (GANs) as an oversampling technique to solve problems with class imbalance.
Generative Adversarial Network - is a class of deep learning framework designed by Ian Goodfellow and his colleagues~\cite{Goodfellow2014}.
The main idea of GANs is a competitive process between two neural networks - a generator and a discriminator.
The generator is trying to deceive the discriminator by producing images similar to the ones from the original dataset.
On the other hand, the main target of the discriminator is to distinguish real data from the generated (fake) one.
As a result, the generator constantly learns how to better fool the discriminator, on the other hand, the discriminator gets constantly better at detecting synthetic images.
A typical GAN architecture is presented in Fig.~\ref{fig:gan_architecture}.

GANs can be described as a way to “unlock” additional information from a dataset.
It has been shown that by oversampling samples that occur with small probability, GANs are able to reduce the false positive rate of anomaly detection~\cite{GanAugmentation}.
This approach is increasingly popular in the medical domain.
For instance, the Vox2Vox method~\cite{Cirillo2021} was designed for 3D segmentation of brain tumours or an augmentation method for CNN-based liver lesion classification~\cite{FridAdar_2018}.
There are works in the literature where GANs were used to augment chest X-ray data samples for COVID-19 detections.
Two of them are focused on the binary classification task, using an AC-GAN-based solution Waheed et al.~\cite{Waheed_2020} obtained an accuracy of $95 \%$ an increase from $85 \%$ in the case of no augmentation.
On the other hand, Motamed et al.~\cite{Motamed_2021} developed the IAGAN model (an architecture based on DAGAN) and achieved a moderate accuracy increase of $2 \%$ points over no augmentation or traditional augmentation approach.
In a work that used a CGAN-based solution to tackle three class problem, Liang et al.~\cite{Liang_2020} obtained an increase of accuracy of $1 \%$ points over a scenario without augmentation.

The goal of this research is to verify if Generative Adversarial Networks-based data augmentation is a suitable method to achieve better performance of medical image data classification in comparison with classical image data augmentation.
A StyleGAN2-ADA~\cite{StyleGAN2_ADA} architecture is selected for GAN-based augmentation as it was designed to be performant for limited data samples.
For this purpose, a study of the multi-class classification efficiency of chest X-ray scans is carried out.
A data sample consisting of normal, COVID-19, viral pneumonia and lung opacity images is used.
To our best knowledge, it is the first try to consistently treat four classes present in COVID-19 chest X-ray datasets with an approach including GAN-based augmentation.
Results are obtained using two state-of-the-art, pre-trained CNN classification models.
The efficiency obtained using classical and GAN-based augmentation is compared to a baseline scenario of no data augmentation.

The manuscript is organised as follows: in the \textit{Introduction} the research problem is described with a review of the literature on the subject, the \textit{Materials and methods} section describes the \textit{Dataset} and its preprocessing followed by \textit{Image comparison metrics} used to assess the quality of the \textit{GAN-based augmentation}.
The section describes, in addition, details of the \textit{Classical augmentation} as well as the approach used for the \textit{Classification} of images.
In the \textit{Results} section classifier output is described with additional discussion of classifier performance based on \textit{Saliency maps}.
The article ends with \textit{Conclusions}.

\section{Materials and methods}

\subsection{Dataset}

\begin{figure}
    \centering
    \subfloat[]{
        \includegraphics[width=0.2\textwidth]{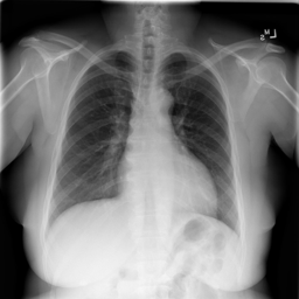}
    }
    \subfloat[]{
        \includegraphics[width=0.2\textwidth]{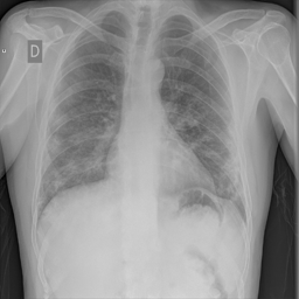}
    }
    \subfloat[]{
        \includegraphics[width=0.2\textwidth]{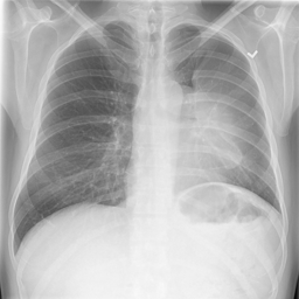}
    }
    \subfloat[]{
        \includegraphics[width=0.2\textwidth]{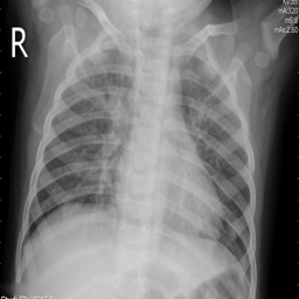}
    }
    \subfloat[]{
        \includegraphics[width=0.2\textwidth]{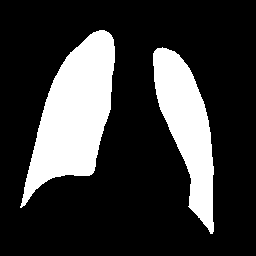}
    }
    \\
    \hfill
    \subfloat[\label{fig:image_crop_f}]{
        \includegraphics[width=0.2\textwidth]{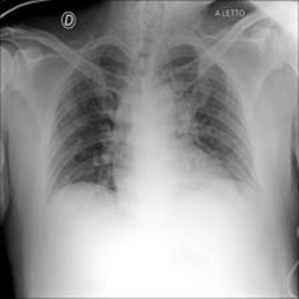}
    }
    \hfill
    \subfloat[\label{fig:image_crop_g}]{
        \includegraphics[width=0.2\textwidth]{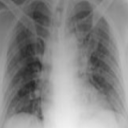}
    }
    \hfill~
    \caption{\label{fig:dataset}Top row - example images from the dataset: a) normal, b) COVID-19 positive, c) lung opacity, d) viral pneumonia, e) lung segmentation mask. Bottom row - Original COVID-19 image (f) and the  same image cropped and resized (g).}
\end{figure}

Research is based on the second update of the "COVID-19 Radiography Database"~\cite{Chowdhury2020,Rahman2021,Covid19Database} developed by a team of researchers from Qatar University, Doha, Qatar, and the University of Dhaka, Bangladesh along with their collaborators from Pakistan and Malaysia in collaboration with medical doctors.
The database incorporates posterior-to-anterior (AP)/anterior-to-posterior (PA) chest X-ray images from multiple public sources~\cite{BIMCV_COVID,Hannover_COVID,SIRM_COVID,cohen2020covid,cohen2020covidProspective,haghanifar2020covidcxnet,RSNA,Kermany2018-ok} and contains:
3616 images of COVID-19-positive cases,
6012 images of lung opacity (non-COVID lung infection),
1345 images of viral pneumonia,
10192 images of healthy lungs.
Images are provided after conversion to common PNG format with 256x256 dimensions.
For each image, the dataset authors provided a corresponding lung segmentation mask obtained using a dedicated U-Net model~\cite{Rahman2021}.
Sample images from the database are presented in Fig.~\ref{fig:dataset}.

Images in the original dataset are not framed consistently, with body parts visible in the image tending to differ depending on the original source.
In order to remove this possible bias all images are cropped to a rectangular region of interest encompassing the corresponding lung mask. After cropping the images are resized to a common size of 128x128 pixels. An example of cropped and rescaled image is presented in Figs.~\ref{fig:image_crop_f} and \ref{fig:image_crop_g}. Finally, the pixel values are normalised to [0;~1] scale. All images are grayscale with three channels present, such format is retained as an input for the classification networks as this is the format they were trained on originally. For the GAN training, a single-channel grayscale format is used which leads to faster execution.

\begin{figure}
    \centering
    \hfill
    \includegraphics[width=0.2\textwidth]{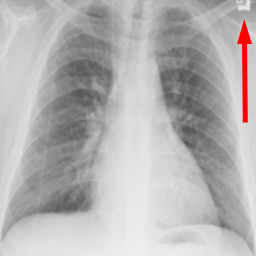}
    \hfill
    \includegraphics[width=0.2\textwidth]{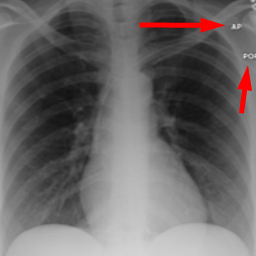}
    \hfill
    \includegraphics[width=0.2\textwidth]{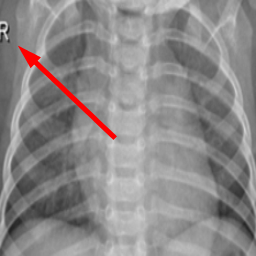}
    \hfill~
    \caption{\label{fig:annotations}Examples of remaining annotations and marks on cropped and resized images.}
\end{figure}

Many images in the dataset contain annotations or marks made during the image acquisition or afterwards by a radiologist.
Examples of such images are presented in Fig.~\ref{fig:annotations}.
Such symbols in the training dataset are known to generate biases as classifiers tend to target their attention to the presence of the symbols instead of genuine features of the image~\cite{Rajaraman_2020,sliency-nature-results}. Therefore after the initial preprocessing, consisting of cropping and resizing all images containing such elements were removed.

The dataset is divided into training, validation and test subsamples. The validation subsample is used to determine hyperparameters of the models e.g. the number of training epochs and for training monitoring. For each class 150 images are selected at random for the validation sample which amounts to $\sim 11\%$ of the smallest class (viral pneumonia). The test subsample is composed of 100 random images from each class corresponding to $\sim 7\%$ of viral pneumonia. It is extended with randomly selected images (200 per class with exception of viral pneumonia where 181 images are available) from a set of images removed during preprocessing due to visible annotation marks. A subsample of just 400 images proved to be too small to provide statistically significant classification performance metrics. The introduction of images with additional features, not present in the training subsample, gives additional insight into the performance of the considered models. It comprises of 1181 images in total and is used to calculate final quality metrics (Table~\ref{tab:results} in \textit{Results}), it is not used during training of the GAN network nor classification networks. The remaining images compose the training subsample consisting of:

\begin{itemize}
  \item 2992 images of COVID-19 positive cases,
  \item 2733 images of lung opacity (non-COVID lung infection),
  \item 1014 images of viral pneumonia,
  \item 7154 images of healthy lungs.
\end{itemize}

The training subsample without any augmentations will be referred to as \textit{no augmentation} scenario throughout the article.

\subsection{Image comparison metrics}

Classic augmentation-generated images contain discriminating features by virtue of being created from a real image.
In the case of synthetic images, obtained from a generative model like a GAN, the quality of the model dictates how well they represent the sample to be augmented.
In order to verify that GAN-generated images are similar to the original sample, several image comparison metrics are considered: Root Mean Square Error (RMSE), Signal to Reconstruction Error Ratio (SRE), Structural Similarity Index Measure (SSIM) and Fréchet Inception Distance (FID). Each metric is briefly described at the end of this section: Eq.~\ref{eq:rmse} for RMSE, Eq.~\ref{eq:sre} for SRE, Eq.~\ref{eq:ssim} for SSIM, and Eq.~\ref{eq:fid} for FID.

\begin{figure}
    \centering
    \subfloat[]{
        \includegraphics[width=0.33\textwidth,trim={9mm 0 12mm 12mm},clip]{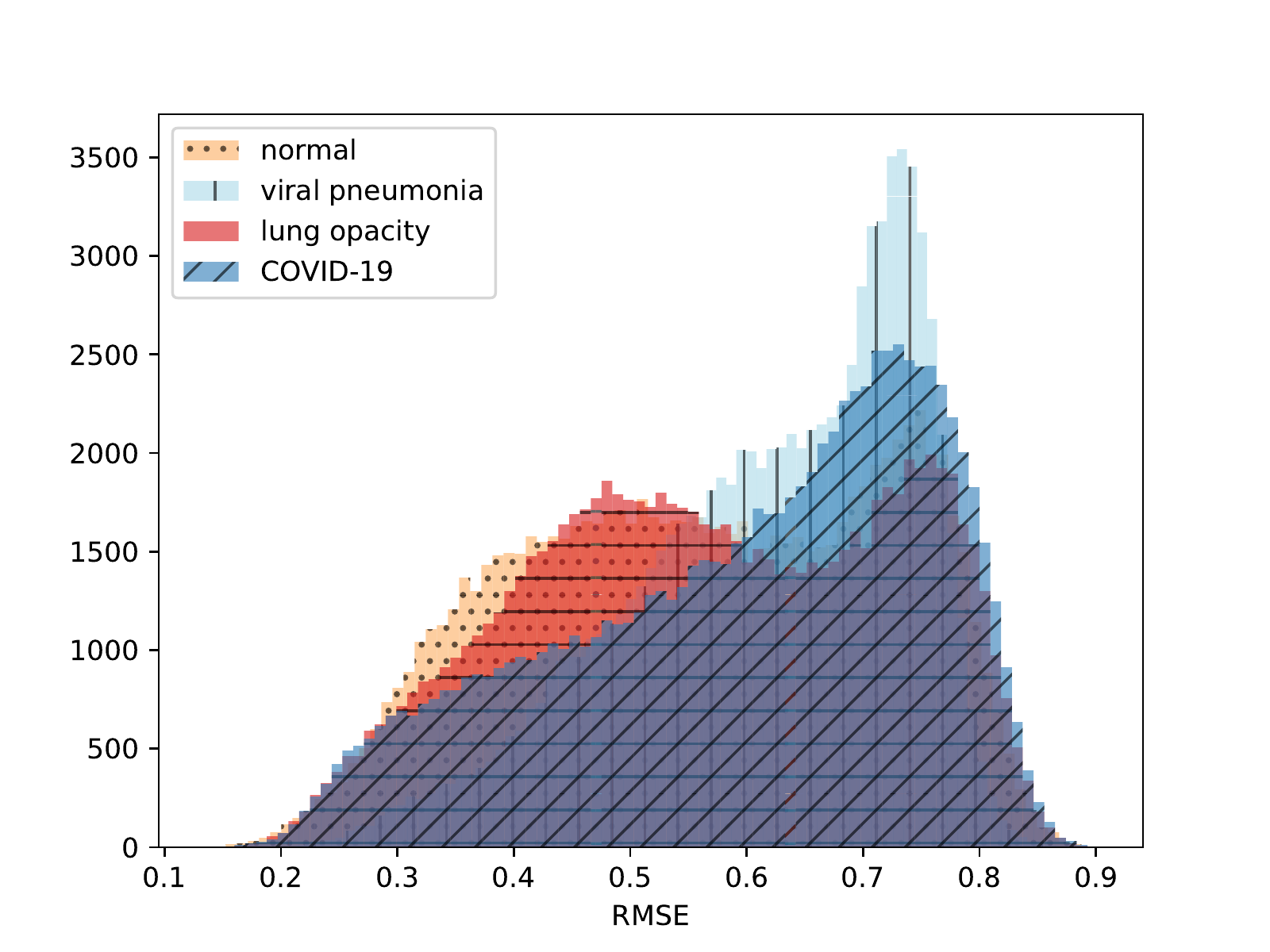}
    }
    \subfloat[]{
        \includegraphics[width=0.33\textwidth,trim={9mm 0 12mm 12mm},clip]{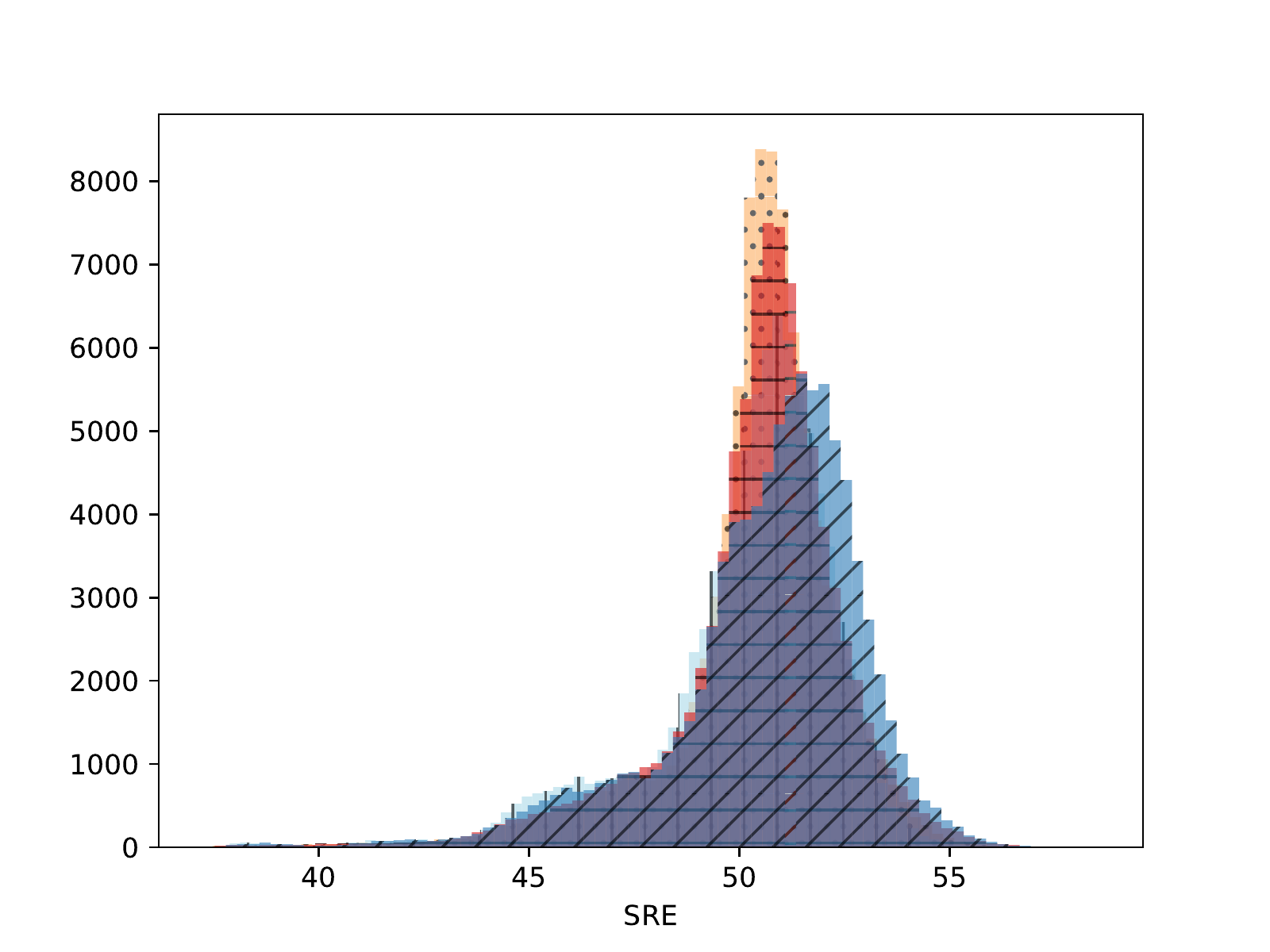}
    }
    \subfloat[]{
        \includegraphics[width=0.33\textwidth,trim={9mm 0 12mm 12mm},clip]{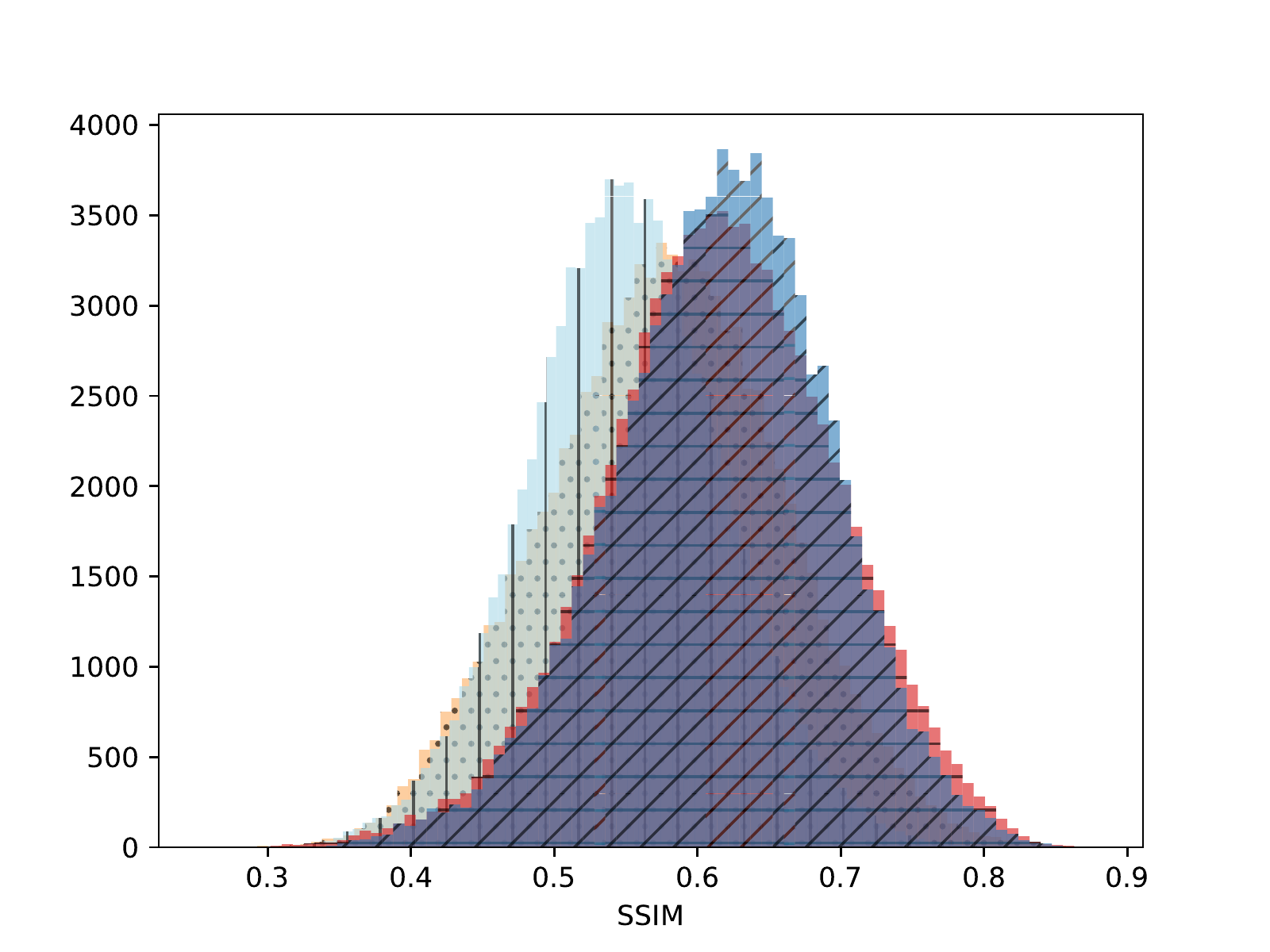}
    }
    \caption{\label{fig:metrics_samples}Distribution of image similarity metrics: a) RMSE, b) SRE and c) SSIM. Four samples are compared to the COVID-19 sample: normal (yellow dotted), viral pneumonia (light blue vertical lines), lung opacity (red horizontal lines) and COVID-19 (blue diagonal lines).}
\end{figure}

To visualise similarity metrics that compare image pairs (RMSE, SRE, SSIM) two metric distributions are generated: intra-similarity and inter-similarity.
For each compared sample 300 random images are selected.
The inter-similarity distribution is calculated for each image pair combination from both compared samples e.g. generated images of COVID-19 and real images of COVID-19.
For the intra-similarity, all pairs of images in a single sample e.g real images of COVID-19 are considered.
By comparing the shape of intra-similarity for the training subsample (\textit{Dataset}) with the inter-similarity of training and generated samples the quality of the generated sample can be judged.
It is also possible to compare in this way between the four image classes present in the dataset.
The resulting distributions of inter-similarity of COVID-19 and the three remaining classes are compared in Fig.~\ref{fig:metrics_samples} to distributions of intra-similarity of the COVID-19 sample for each metric.
All metrics, as expected based on available medical evidence~\cite{Rubin2020-nd}, indicate that there are visible differences not only between COVID-19 and normal samples but also between viral pneumonia and lung opacity.
Therefore, all classes should be distinguishable.

\begin{figure}
    \centering
    \subfloat[FID $\approx 3420$]{
        \includegraphics[width=0.2\textwidth]{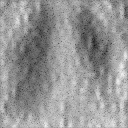}
    }
    \hfill
    \subfloat[FID $\approx 737$]{
        \includegraphics[width=0.2\textwidth]{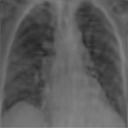}
    }
    \hfill
    \subfloat[FID $\approx 338$]{
        \includegraphics[width=0.2\textwidth]{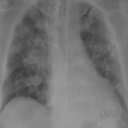}
    }
    \caption{\label{fig:fid_generated_images}Examples of GAN-generated images at different stages of the training process corresponding to decreasing FID values.}
\end{figure}

In contrast to classical metrics, the FID compares whole datasets not individual images, as described in \textit{Fréchet Inception Distance} section.
The sensitivity of this metric to image distortions is visualised in Fig.~\ref{fig:fid_generated_images} using GAN-generated images from different stages of training (epochs: 16, 96 and 872).

\subsubsection{Classical metrics}

The RMSE is one of the common metrics available. A square root of the sum of deviations for each pixel and channel is calculated as (Eq.~\ref{eq:rmse}):
\begin{equation}\label{eq:rmse}
RMSE(\bm{x},\bm{y}) = \sqrt{\frac{1}{N}\sum_{i=1}^{N} (x_{i} - y_{i})^2}.
\end{equation}
An RMSE equal to 0 indicates the perfect identity of compared images.
The maximal value depends on the dynamic range of samples.
While it is fast to calculate it is susceptible to noise and is easily biased by hot regions.

The SRE~\cite{SRE} expresses the ratio between the signal power and the power of distorting noise.
It is very similar to the Peak signal-to-noise ratio (PSNR) however measures error relative to the mean image intensity, hence is better suited for images of varying brightness.
In image processing one usually treats an original image as the signal while the noise is the difference from a noisy representation e.g. a compressed image.
As signals can have a very wide dynamic range, the SRE is usually expressed in a logarithmic decibel scale.
For 2D images when using  the Frobenius norm it can be expressed as (Eq.~\ref{eq:sre}):
\begin{equation}\label{eq:sre}
SRE(\bm{x},\bm{y}) = 10 log_{10}\frac{\mu_{\bm{x}}^2}{RMSE^2(\bm{x},\bm{y})},
\end{equation}
where the $RMSE$ is Root Mean Squared Error of two images and $\mu_{\bm{x}}$ is is the average value of $\bm{x}$.
The higher the SRE, the higher the similarity of compared images, for completely unrelated images it approaches 0.
For two identical images, the SRE will be undefined as $RMSE$ present in the denominator is equal to zero.

Most classical methods of image comparison (like RMSE or SRE) attempted to quantify the visibility of differences between a distorted image and a reference one.
The SSIM~\cite{SSIM} is designed under the assumption that human visual perception is adapted for extracting structural information.
Therefore it quantifies the image differences based on the degradation of structural information.
The metric is based on three comparison measurements between the samples of $x$ and $y$: luminance, contrast and structure.
The luminance difference is modelled using mean pixel values, the contrast by pixel variance and structure with the covariance of two samples.
The three measures can be reduced to a compact form (Eq.~\ref{eq:ssim}):
\begin{equation}\label{eq:ssim}
    SSIM(\bm{x}, \bm{y}) = \frac{(2 \mu_{\bm{x}} \mu_{\bm{y}} + c_1)(2 \sigma_{\bm{xy}} + c_2 ) }{ ( \mu^2_{\bm{x}} + \mu^2_{\bm{y}} + c_1 ) ( \sigma^2_{\bm{x}} + \sigma^2_{\bm{y}} + c_2 ) },
\end{equation}
with $\mu_{\bm{i}}$ being the average of the image $\bm{i}$, $\sigma_{\bm{i}}^{2}$ the variance of the sample $\bm{i}$, $\sigma_{\bm{ij}}$ the covariance of the two samples, $c_{1}$ and $c_{2}$ two constants to stabilise the division with weak denominator.

The value of the SSIM metric falls between -1 and 1, where 1 indicates perfect similarity, 0 indicates no similarity, and -1 indicates perfect anti-correlation.
As image statistical features and possible image distortions are position dependant it is useful to apply SSIM locally rather than globally.
The process is realised by using a sliding window of fixed size.
This method is often referred to as the Mean Structural Similarity Index (MSSIM).

\begin{figure}
    \centering
    \includegraphics[width=0.3\textwidth]{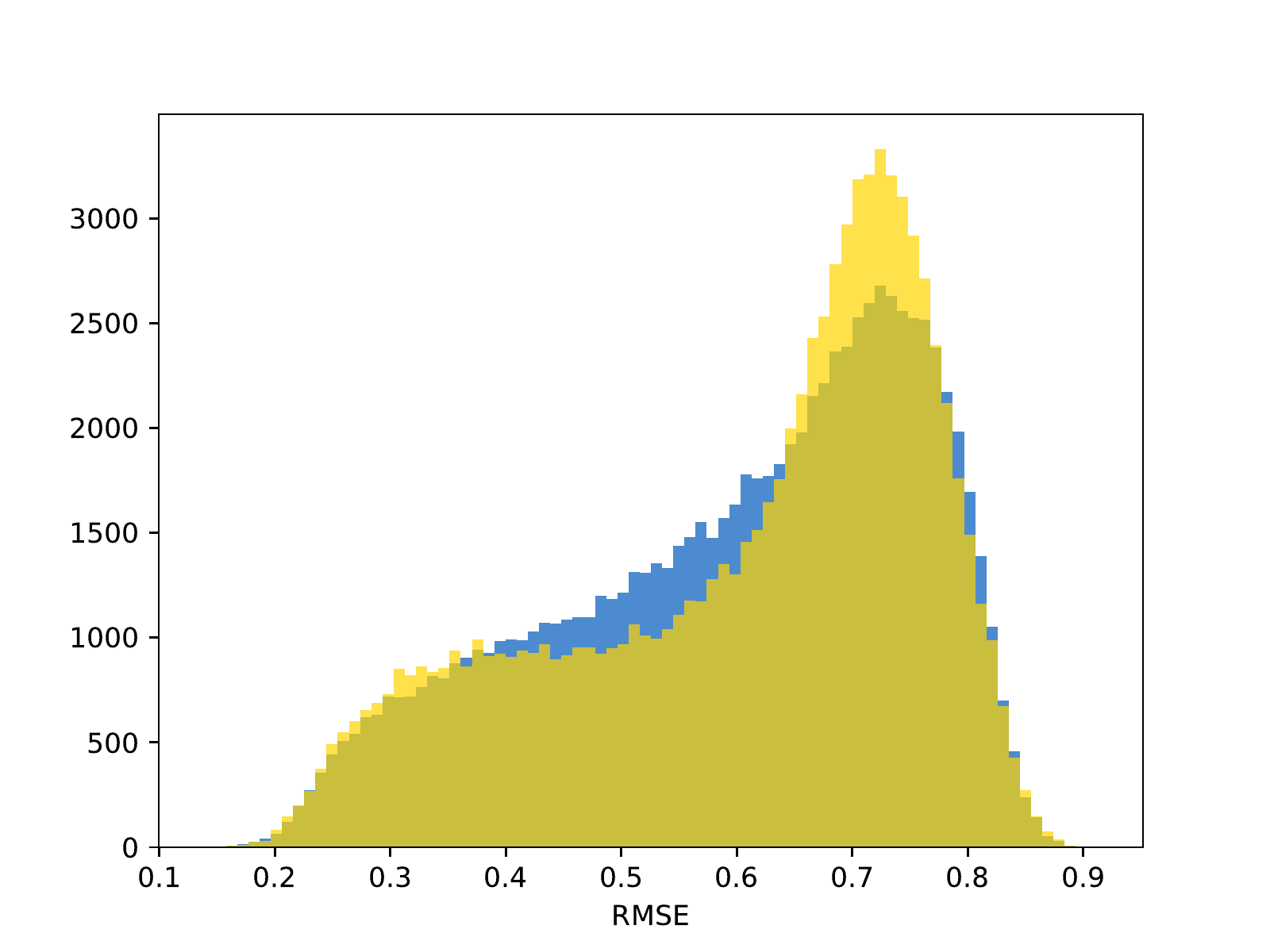}
    \includegraphics[width=0.3\textwidth]{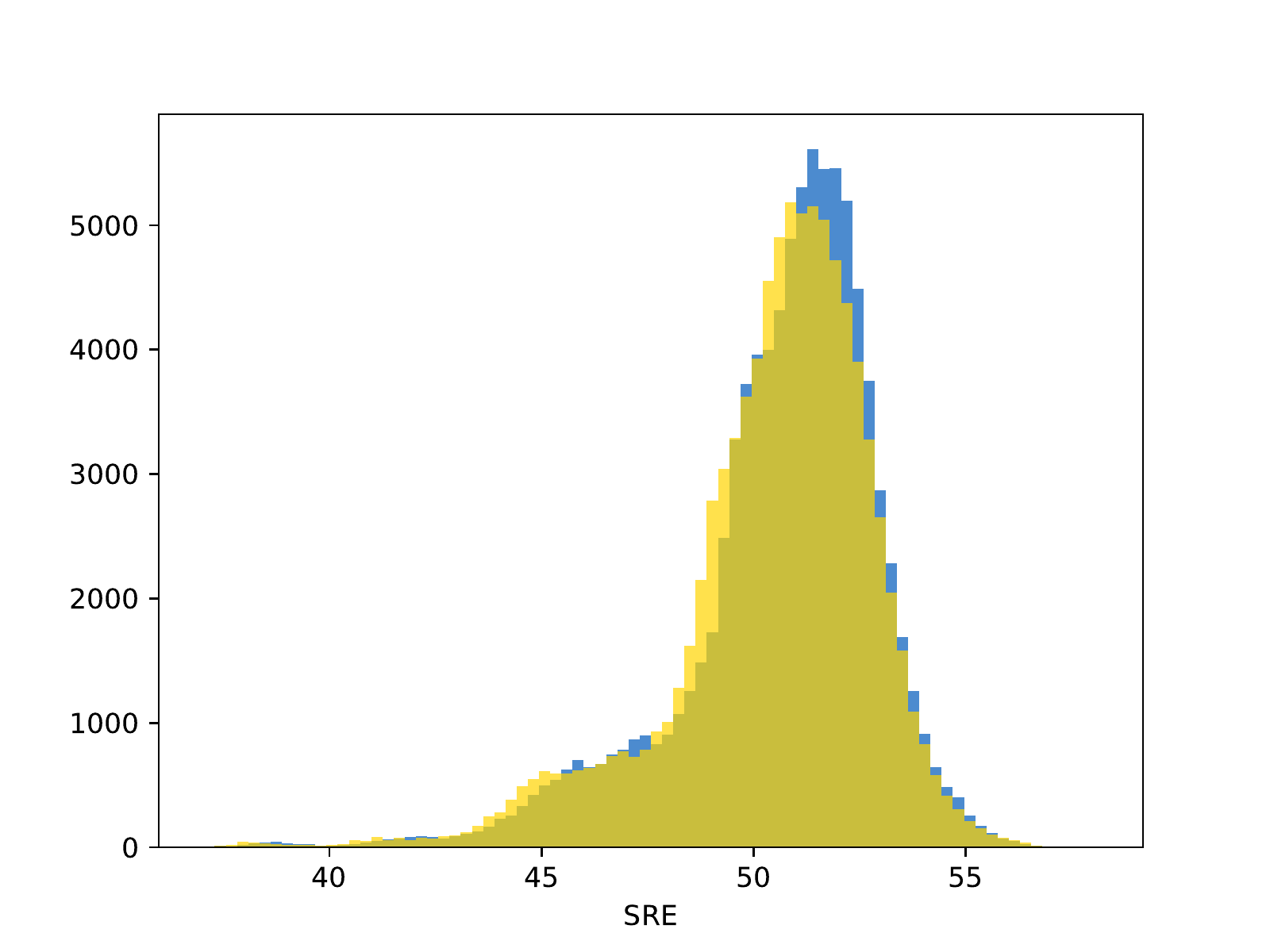}
    \includegraphics[width=0.3\textwidth]{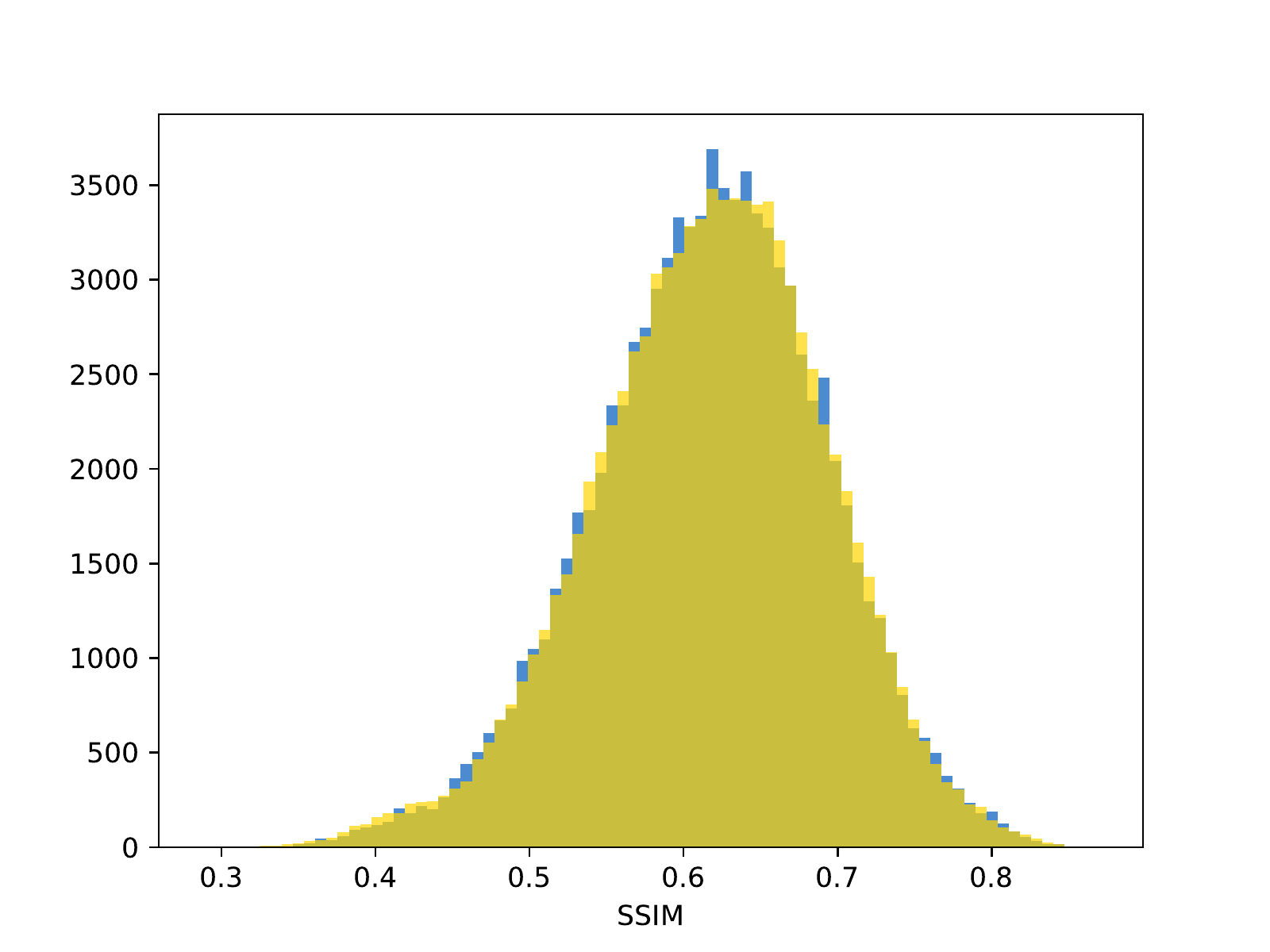}
    \\
    \includegraphics[width=0.3\textwidth]{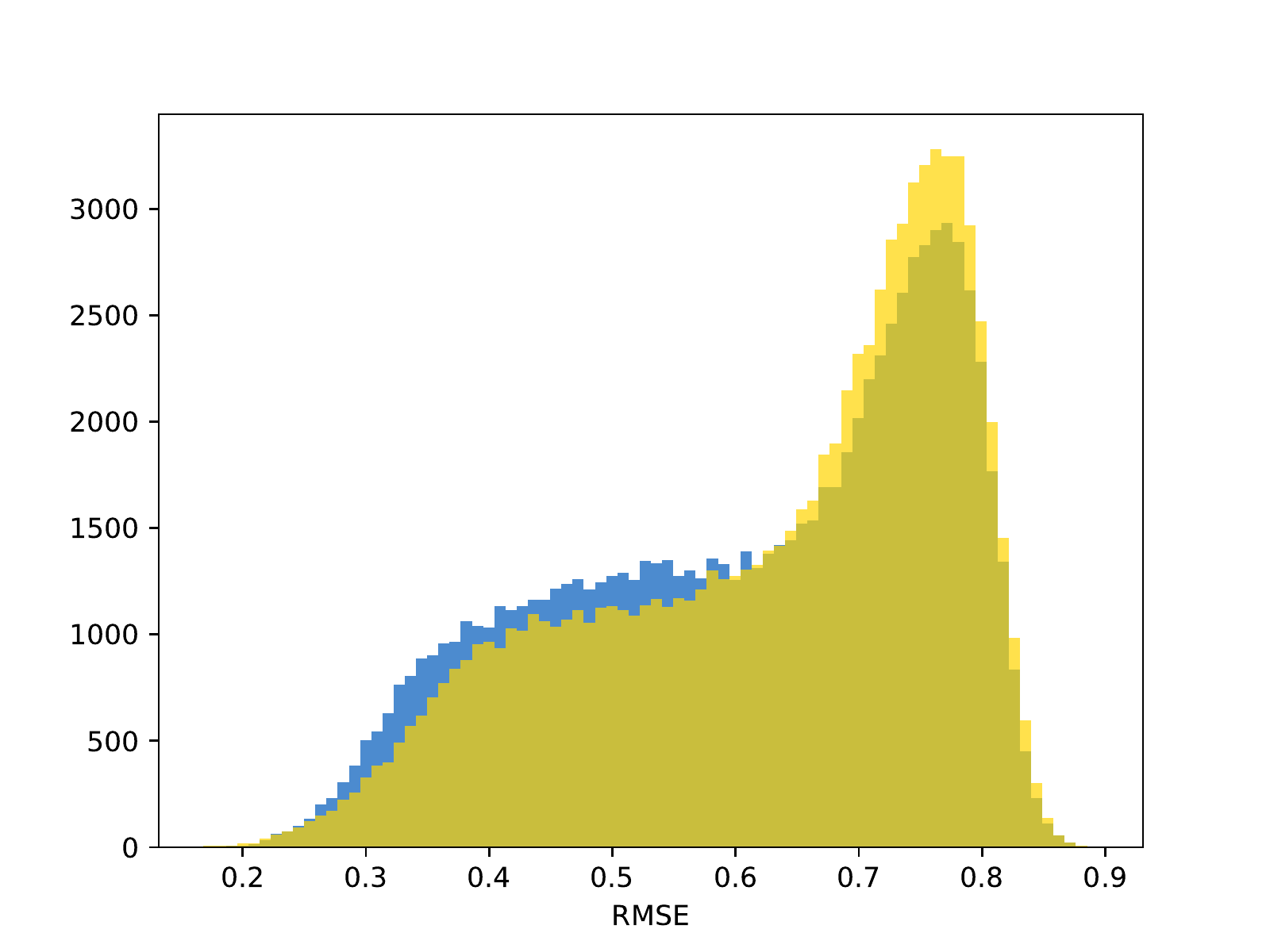}
    \includegraphics[width=0.3\textwidth]{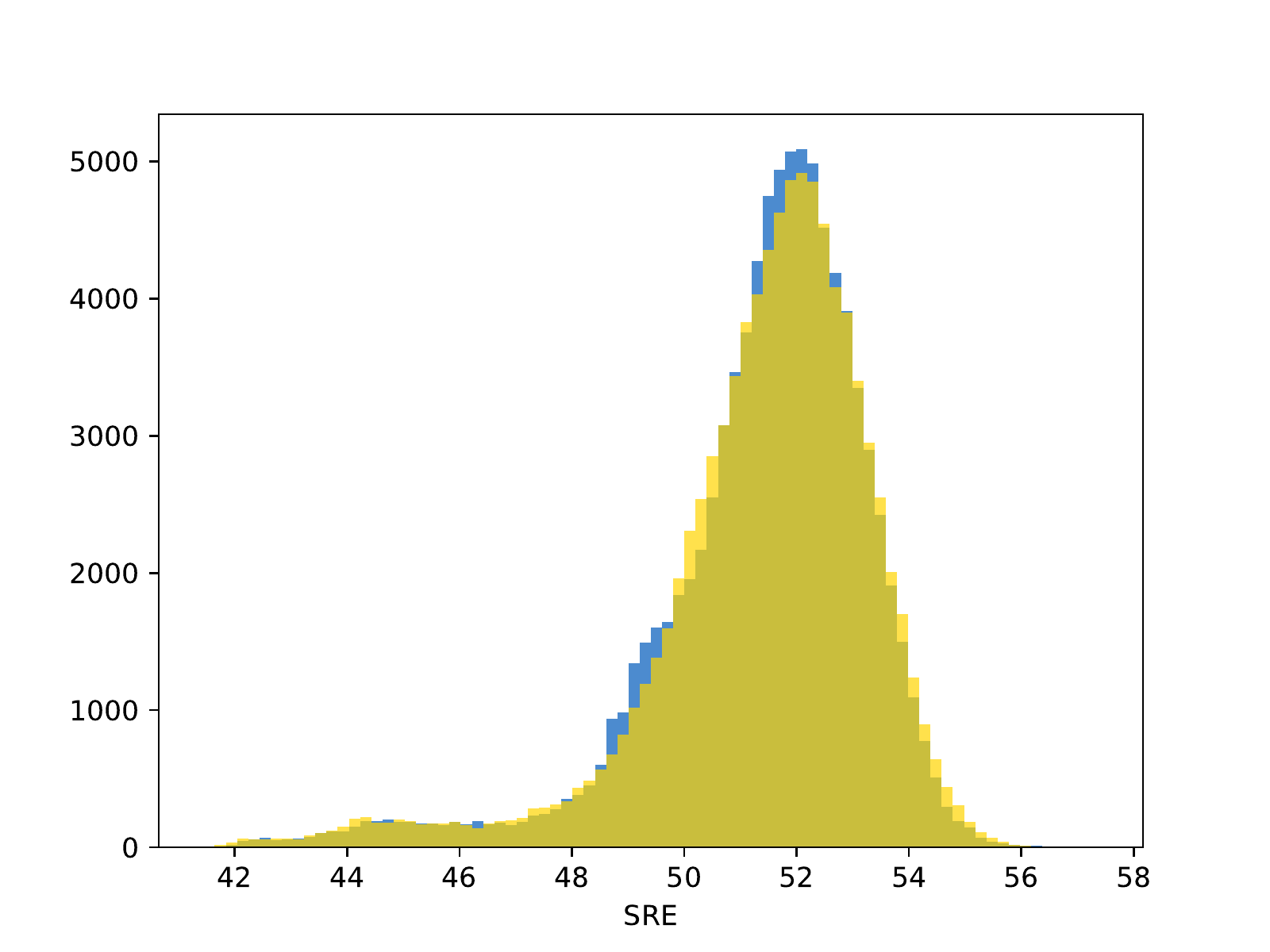}
    \includegraphics[width=0.3\textwidth]{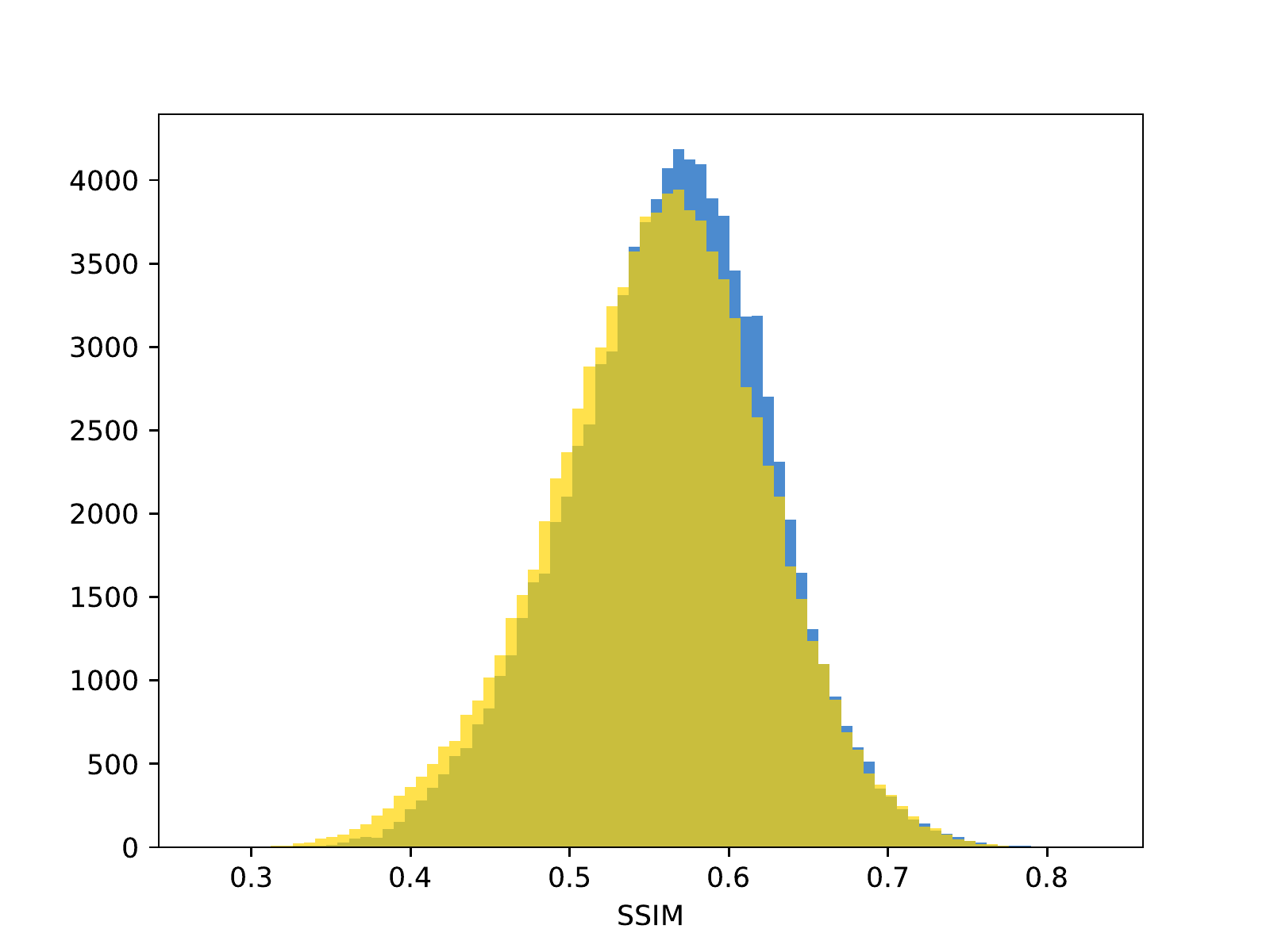}
    \\
    \includegraphics[width=0.3\textwidth]{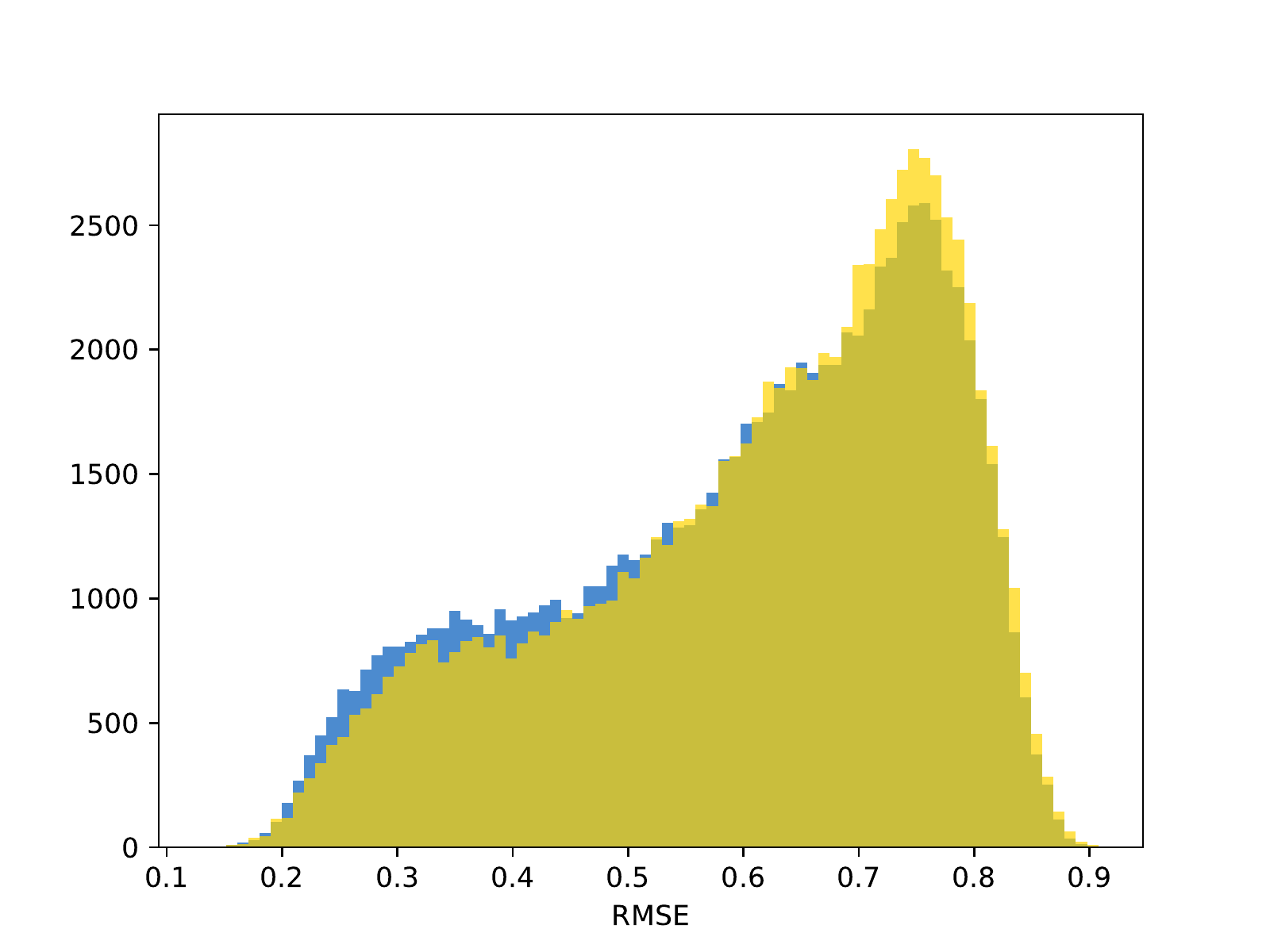}
    \includegraphics[width=0.3\textwidth]{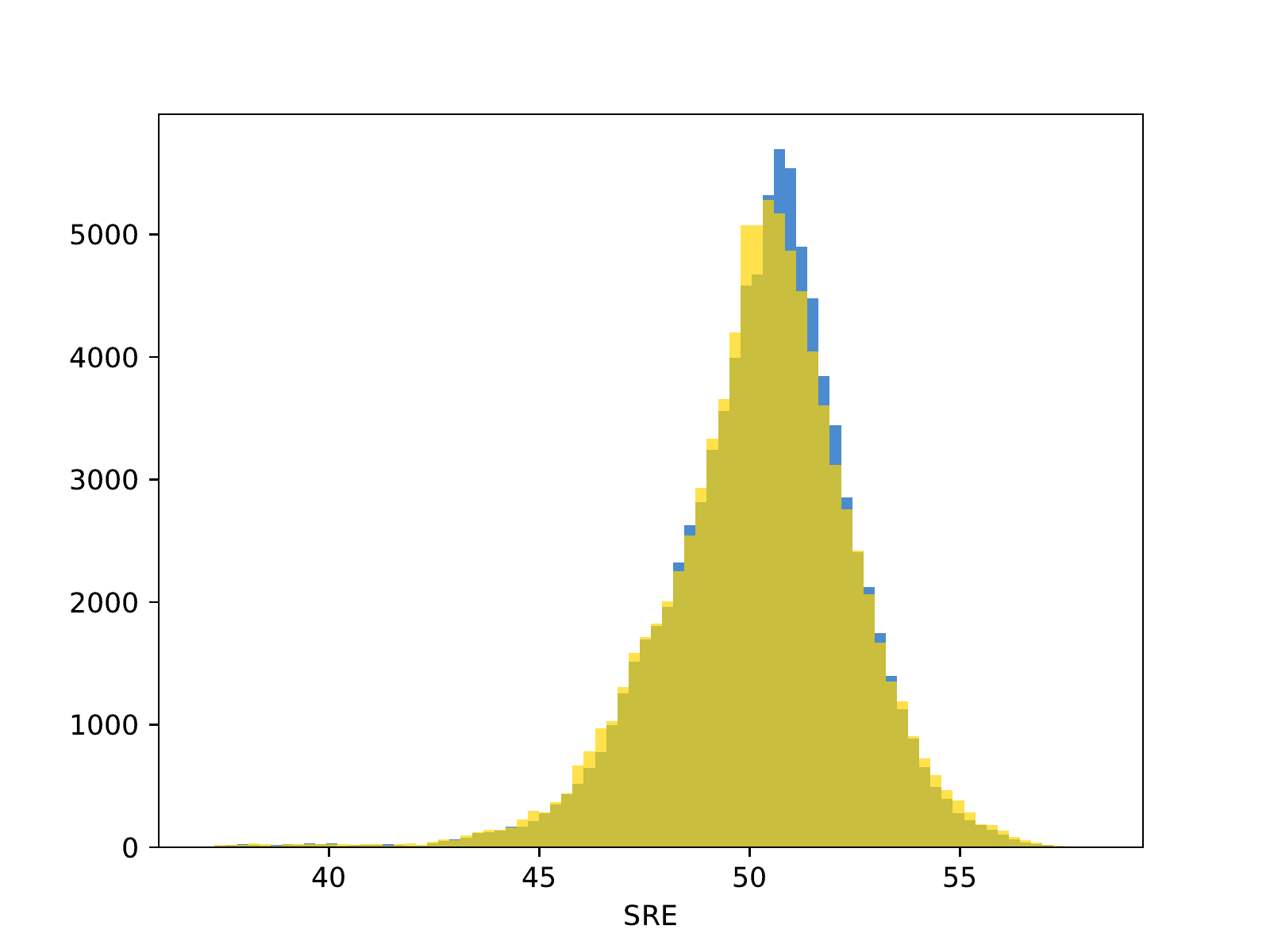}
    \includegraphics[width=0.3\textwidth]{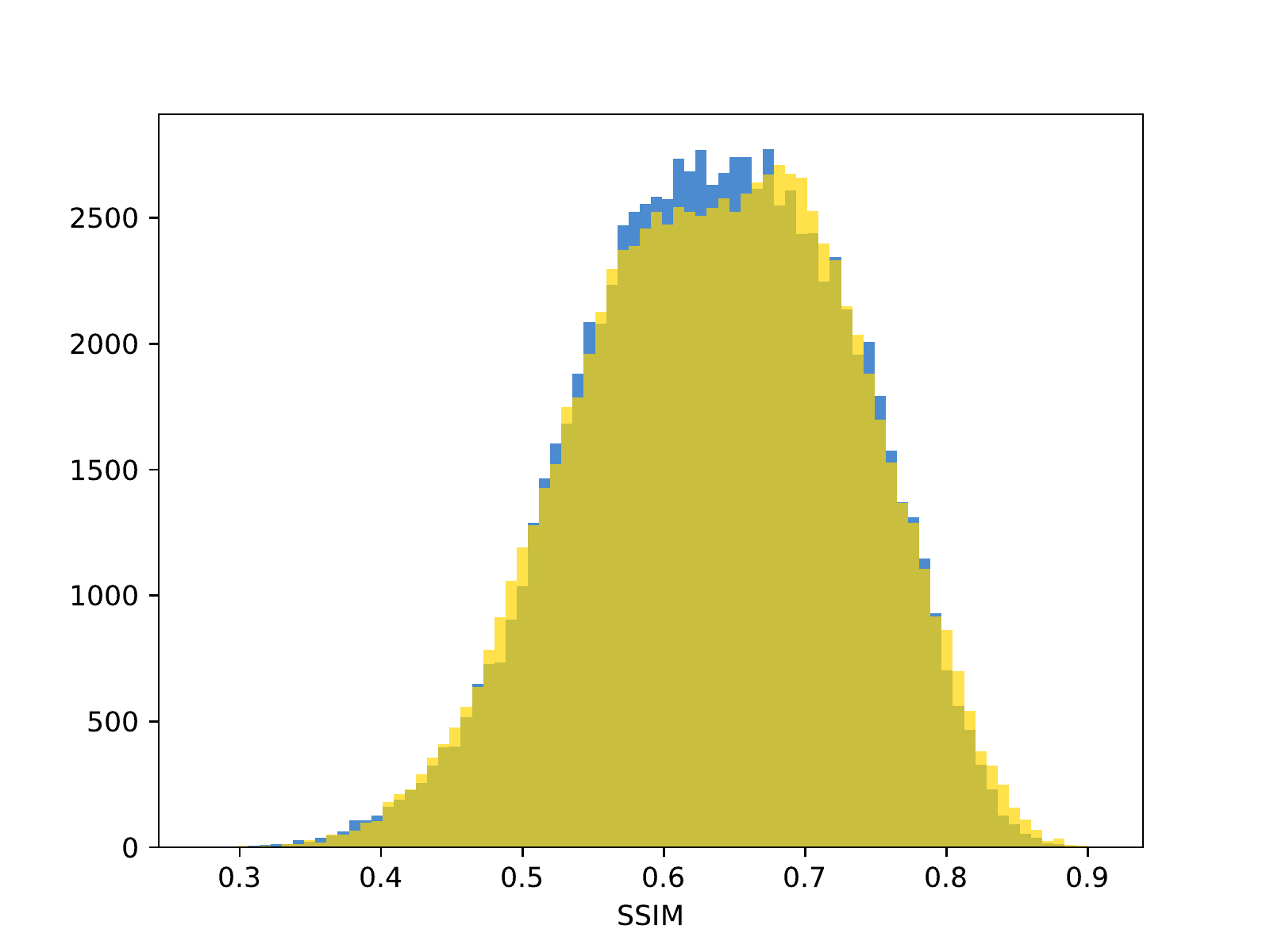}
    \caption{\label{fig:metrics_generated}Distribution of image similarity metrics: RMSE (left column), SRE (central column) and SSIM (right column). For three samples GAN generated images are compared to real ones: COVID-19 (top row), viral pneumonia (central row) and lung opacity (bottom row). The blue distribution corresponds to the intra-similarity of images in the real sample and the yellow distribution corresponds to the inter-similarity between generated and real samples.}
\end{figure}

To visualise differences between synthetic and real samples distributions are generated for each of the RMSE, SRE and SSIM metrics: intra-similarity and inter-similarity.
For each compared sample 300 random images are selected.
The inter-similarity distribution is calculated for each image pair combination from both compared samples e.g. generated images of COVID-19 and real images of COVID-19.
For the intra-similarity, all pairs of images in a single sample e.g real images of COVID-19 are considered.
The distributions for generated samples describe very well the distributions for corresponding real images as shown in Fig.~\ref{fig:metrics_generated}.

\subsubsection{Fréchet Inception Distance}

Most classical image comparison metrics are designed to quantify distortion of the original image by compression or reconstruction algorithms.
They are not well suited as similarity measures of different images of the same class or corresponding to several classes.
The main drawback is that the spatial localisation of features in the images plays a crucial role in all of those metrics.
Comparing two pictures of a cat in different poses will result in metrics indicating substantial differences while the underlying object is the same.
This aspect of the classical metrics is problematic when evaluating the quality of synthetic images from generative algorithms.
It would be beneficial to asses the reproduction of a given class as a whole and not the identity with particular images.
An attempt to solve this problem is the Fréchet Inception Distance~\cite{FID}.
Images are fed into the Inception-v3 model in order to obtain vision-relevant features.
The labels in the model output vector can be treated as dimensions, hence for a sample of images, one obtains a multidimensional distribution.
Images with meaningful objects should belong to a few object classes and hence have low label entropy.
On the other hand for the large variance of images in a sample, as expected from GAN-generated output, the label entropy should be high across images.
For synthetic images that properly mimic features of the original sample, those distributions should be similar.
In order to simplify the comparison one only considers the distribution of the first two moments: mean $\bm{m}$ and covariance $\bm{C}$.
Then the distributions can be approximated by multidimensional Gaussian distributions.
With those assumptions, a Fréchet distance $d$~\cite{Frechet_1957,Dowson_1982} of the label distributions for two samples can be defined as in Eq.~\ref{eq:fid}.
Thus to assess the similarity of a synthetic image sample to the original one the FID value can be calculated.
This metric approaches zero for identical samples.
\begin{equation}\label{eq:fid}
    d^2((\bm{m}_x,\bm{C}_x),(\bm{m}_y,\bm{C}_y))=||\bm{m}_x-\bm{m}_y||^2_2+Tr(\bm{C}_x+\bm{C}_y-2(\bm{C}_x\bm{C}_y)^{1/2})
\end{equation}

\subsection{Classical augmentation}

Classical image data augmentation can be defined as algorithms for the generation of synthetic images from a real data sample that are not based on machine learning~\cite{AsurveyonImageDataAugmentation}. In the case of data augmentation in computer vision following basic operations on images are usually considered:
\begin{itemize}
    \item kernel filters - used in image convolution to obtain sharpening or blurring;
    \item random erasing - randomly set image regions to the specified value;
    \item colour space transformations - image darkening, brightening, changing the value of individual colour channels or operations on an image histogram;
    \item geometric transformations - image affine transformations or simulation of lens distortions;
    \item mixing images - addition or subtraction of whole images or image regions between random image pairs.
\end{itemize}

\begin{figure}
    \centering
    \includegraphics[width=0.2\textwidth]{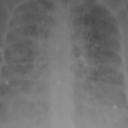}
    \includegraphics[width=0.2\textwidth]{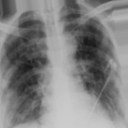}
    \includegraphics[width=0.2\textwidth]{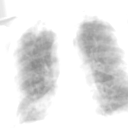}
    \includegraphics[width=0.2\textwidth]{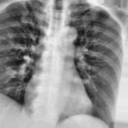}
    \caption{\label{fig:augmented_images}Examples of chest X-ray images after application of classic augmentation transformations.}
\end{figure}

Selecting a set of the best operations for a particular task is not obvious and there is no clear guidance on how to proceed.
Excessive augmentation will disturb the results, while too weak will not increase the sample variance hence not preventing overfitting.
In order to have a baseline for GAN-based augmentation a set of typical operations for image augmentation is selected that maximises accuracy on the validation subsample (\textit{Dataset}) using  EfficientNet-B0 architecture.
As a result following image transforms and their parameters are used for the classical augmentation scenario: 
\begin{itemize}
    \item rotation -  randomly rotate the image by an angle of up to 5 degrees clockwise or counterclockwise;
    \item shift - randomly shift the image along cardinal axes within the range of 5$\%$ of the specific image size, the empty field is filled with the trace of the last shifted pixels;
    \item stretch - randomly stretch the image between opposite vertices by up to 5 $\%$;
    \item zoom - randomly zoom inside pictures up to 15$\%$ of the specific value of the image size;
    \item brightness change - randomly enlighten or darken the image by up to $40\%$.
\end{itemize}
The parameters of image processing are selected by a grid search with the following steps: rotation, shift and stretch adjusted jointly with values 0, 5 and 10; zoom with values 10, 15 and 20; brightness change with values 30, 40 and 50$\%$. This yields 27 combinations, from which the one yielding the highest accuracy on the validation subsample (\textit{Dataset}) is selected.
Synthetic images are generated using such a classical augmentation approach for three underrepresented samples (COVID-19, viral pneumonia and lung opacity) to equalise their size with the normal sample.
Example results of such classical augmentation are presented in Fig.~\ref{fig:augmented_images}.
The resulting augmented data sample will be referred to as \textit{classical augmentation} scenario throughout the article.

\subsection{GAN-based augmentation}

In this work, we propose to generate synthetic images for underrepresented classes using the StyleGAN2-ADA architecture~\cite{StyleGAN2_ADA}.
The StyleGAN2-ADA model was designed to increase training stability on limited datasets using adaptive augmentation while being robust for high-resolution images.
In addition, it can be extended to offer class-conditional image generation.
By training a single GAN model for all underrepresented classes one effectively increases the training sample in case there are common features between classes.

While first GAN architectures were able to learn to generate relatively small images, further research allowed to overcome this limitation.
In particular progressive growing of GAN convolutional layers proved to be very successful~\cite{PG_GAN}.
In December 2019 an NVIDIA team has introduced a better version of GAN architecture - StyleGAN~\cite{StyleGAN}.
The StyleGAN departs from traditional GAN architecture as for the input to the generator a learned constant is used instead of a latent code.
The latent code $\bm{z}$ is injected into each generator block through an additional non-linear mapping network $f: \mathbb{Z} \to \mathbb{W}$.
With given latent code $\bm{z}$ in latent space $\mathbb{Z}$, this network produces $\bm{w} \in \mathbb{W}$.
This architecture is able to automatically separate high-level image attributes such as human pose, gender, and stochastic variation such as hair style and colour, wrinkles, etc.
Depending on the depth at which latent code mapping is introduced it controls coarse or fine style features of the image.
This allows for style mixing when different latent codes are provided to separate generator blocks.
In addition, authors introduced techniques such as adaptive instance normalisation and weight demodulation that improved the quality of the generation process.

Later, in March 2020 StyleGAN authors introduced several changes to the original StyleGAN architecture that improved the quality of generated images~\cite{StyleGAN2}.
In particular, the generator normalisation was redesigned and progressive growing was made more robust by introducing skip connections. In June 2020 an adaptive discriminator augmentation mechanism was introduced that significantly stabilises training with limited data leading to StyleGAN2-ADA architecture~\cite{StyleGAN2_ADA}.

Adaptive discriminator augmentation (ADA) is of particular interest.
In order to improve GAN stability for small datasets the StyleGAN architecture was modified to incorporate an augmentation step for both the generator and discriminator.
The augmentations used are traditional image transforms e.g. rotations or color transfromations.
The ADA mechanism solves the problem of augmentation leaks.
An augmentation leak appears when the GAN generator learns the augmented data distribution instead of just the original one.
For example, if applying random image rotation augmentation with the equal probability it is impossible to determine the orientation of the original image.
Adaptive discriminator augmentation solves this problem by applying augmentations with adaptively changing their probability. 
In order to tell if the discriminator is overfitting a heuristic $r_t$ is used that estimates the portion of the training set that gets
positive discriminator outputs:

\begin{equation}\label{eq:ada_rt}
    r_t = \mathbb{E}[sign(D_{train})]
\end{equation}

Where $D_{train}$ denotes the discriminator output for the training set while $\mathbb{E}$ indicates a mean over N consecutive mini-batches.
The $r=0$ means no overfitting and $r=1$ means complete overfitting for validation and test dataset respectively.
The probability $p$ of a given augmentation is used to control its strength.
Initialised as $p=0$, it is adjusted during training based on the heuristics described above.
If the heuristic's value indicates too much or too little overfitting $p$ is being increased or decreased by a fixed number.
In that way, it is possible to effectively control the augmentation to prevent leaks and be able to train GAN with smaller datasets.
One very interesting extension of StyleGAN2-ADA architecture proposed by its authors is class-conditional image generation.
For the generator, the class identifier is embedded into a 512-dimensional vector that is concatenated with the original latent code after normalisation.
For the discriminator, the output is weighted with the embedded class identifier.

In the presented analysis the official PyTorch StyleGAN2-ADA implementation is used with class-conditional generation enabled.
A single model is trained to generate images depending on the selected class label.
In the proposed GAN augmentation approach the synthetic images are generated only for underrepresented classes in order to obtain a fully balanced dataset.
Therefore the inclusion of the normal (healthy) class of images is not necessary for the GAN training.
Nevertheless, two scenarios are analysed to verify if providing additional images with similar overall spatial features is beneficial: GAN training on 3 classes of images and on 4 classes of images.
In both cases training time is observed to be similar.

The target for $r_t$ ADA heuristic (Eq.~\ref{eq:ada_rt}) is set to 0.6. 
Both generator and discriminator learning rates are set to 0.0025 while the batch size is set to 32.
Following StyleGAN2 authors~\cite{StyleGAN2}, we use non-saturating logistic loss with $R_1$ regularisation with $gamma$ set to 1.024.
All other parameters are set to default values provided by the NVIDIA implementation~\cite{StyleGAN2_ADA}.

To increase the training speed all images from the training subsample  (\textit{Dataset}) are converted to grayscale single-channel representation.
After such optimisation, the average training run on NVIDIA Tesla K80 takes around 53 hours while processing around 876,000 images.

To eliminate overfitting, image quality is monitored throughout the training process.
The Fréchet Inception Distance is calculated every epoch for each image class independently between generated and real data samples.
It is evaluated on the training subsample (\textit{Dataset}) as the FID metric is known to be biased for small datasets~\cite{StyleGAN}.
In addition, the quality of generated images is verified by calculating RMSE, SRE and SSIM metrics.
The training process is stopped when discriminator overfitting appears and the quality of generated images starts to decrease.
The epoch that corresponds to the smallest mean FID metric from all image classes is selected as a base for the GAN augmentation network.

A model trained with 3 classes is selected for GAN-based augmentation as it results in better image quality as described in \textit{Results}.
It is used to generate images for the three underrepresented samples (COVID-19, viral pneumonia and lung opacity) to obtain a perfectly balanced dataset.
The resulting augmented data sample will be referred to as \textit{GAN augmentation} scenario throughout the article.

\subsection{Classification}

Two popular classification network architectures are used: Inception-v3 and EfficientNet-B0. The former is an image recognition model developed by Christian Szegedy, et. al. in 2015~\cite{Inception} that has been shown to attain greater than $78.1\%$ accuracy on the ImageNet dataset. The latter introduced by Tan et al. in 2019~\cite{EfficientNet} is a convolutional neural network architecture that scales across all of the dimensions using the compound coefficient method.

The fundamental idea behind the Inception type architecture is an inception block.
Instead of passing data through one layer it takes the previous layer input and passes it through four different operations in parallel and then concatenates the outputs from all these different layers.
To combat the vanishing gradient problem in very deep networks Inception-v3 uses auxiliary classifiers.
In addition several optimisations are introduced, in particular, factorisation into smaller convolutions which allows reducing the computational cost.
Furthermore, an efficient grid size reduction is proposed where convolution and pooling are carried out in parallel blocks of stride 2 with results concatenated.

EfficientNets convolutional neural networks were predominantly designed with specified dimensions for particular resource limits.
Having more computational resources model accuracy was often improved by increasing separately the number of layers or number of neurons in a layer (layers width)~\cite{ResNetScaling} or less often input image resolution.
Proper selection of neural network dimensions resulting in accuracy improvement is non-trivial and requires exploration of additional hyperparameters.
In EfficientNets model dimensions can be matched to specific resources by using the compound coefficient technique.
This scaling method defines depth~($d$), width~($w$) and input image resolution~($r$) through constants scaled by a coefficient $\phi$: $d=\alpha^\phi; w=\beta^\phi; r=\gamma^\phi$.
Scaling without a strong baseline model would be ineffective, hence main EfficientNets building blocks called MBconv are based on the MobileNetV2 model.
MobileNetV2 is an efficient architecture designed for environments with constrained resources.
Based on the observation that the computational cost of regular convolution is proportional to $d$, $w^2$ and $r^2$, authors constrained scaling coefficients to $\alpha \cdot \beta^2 \cdot \gamma^2 \approx 2$ such that the total computational cost will approximately increase by $2^\phi$.

The baseline model from EfficientNets is called EfficientNet-B0 from which different versions are scaled using different coefficients up to the most demanding EfficientNet-B7 architecture.
In particular, EfficientNet-B7 achieves state-of-the-art 84.3\% top-1 accuracy on ImageNet, while being 8.4x smaller and 6.1x faster on inference than the best existing convolutional network at the time it was published \cite{EfficientNet}.
As the task of interest is not very complicated, with relatively small image dimensions and four classes to consider, a baseline neural network model is considered to prevent overfitting to the training data. The EfficientNet-B0 was selected as a baseline architecture from one of the best classification convolutional neural networks available.

The default built-in Keras~\cite{Keras} implementations of Inception-v3 and EfficientNet-B0 models are used.
Transfer learning is used starting with library-provided ImageNet~\cite{ImageNet} weights.
Both of the networks are used with their default heads replaced with a custom set of output layers:
\begin{itemize}
    \item Flatten layer
    \item Dense layer with 10 neurons and exponential linear unit (ELU) activation function
    \item Dense layer with 20 neurons and ELU activation function
    \item Dense layer with 4 neurons and softmax activation function
\end{itemize}
Categorical cross-entropy is picked as the loss function and the learning rate is set to Keras default of 0.001.
Optimizer is selected independently for the two networks based on accuracy obtained on the validation subsample (\textit{Dataset}).
In the case of EfficientNet-B0, the Adam optimizer is used while for the Inception-v3 model, the RMSprop is selected.

Using the preprocessed images (as described in the \textit{Dataset} section) the two CNN architectures are trained using transfer learning with four different scenarios:
\begin{itemize}
    \item \textit{no augmentation} with the raw, unbalanced training subsample (\textit{Dataset});
    \item \textit{balanced dataset} with random duplication of images in underrepresented classes from training subsample (\textit{Dataset});
    \item \textit{classical augmentation} corresponding to training subsample (\textit{Dataset}) after balancing procedure described in \textit{Classical augmentation};
    \item \textit{GAN augmentation} utilising training subsample (\textit{Dataset}) equalised as described in \textit{GAN-based augmentation}.
\end{itemize}
Apart from the first scenario the size of underrepresented samples is increased, using the relevant technique, to be equal to the normal sample.
For overfitting monitoring the same validation subsample (\textit{Dataset}), without any augmentation, is used for each model and all scenarios.
All networks were trained with 5 runs of 25 epochs (with batch size equal to 32).
The number of epochs was selected to be equal for each scenario and corresponds to the point where the training accuracy reaches a plateau while validation accuracy does not indicate overfitting of the models.
Training run with the best validation accuracy was picked as the final result. 
The overall quality of each augmentation approach is estimated on the test subsample (\textit{Dataset}) of 1181 images (Table~\ref{tab:results} in \textit{Results}).

\subsection{Saliency maps}
Saliency maps assign pixels in an image according to their contribution (a class score derivative) to the final output of a CNN-based classifier.
Another interpretation of saliency maps is that the magnitude of the derivative indicates which pixels need to be changed the least to affect the class score the most~\cite{saliency-map-base}.
A saliency map is extracted for a given image and class label by back-propagation procedure.
This technique helps extract information from the black box model and visualise how it interprets the given data.
Saliency maps are one of the essential tools that allow verifying that classification results do not depend on artefacts from the training set, so-called "shortcuts", and help in understanding where observed misclassifications come from~\cite{saliency-why}. Saliency maps for instance can be utilised to identify whether the classification of COVID-19 is based on the infected regions of the lungs, rather than being influenced by image artefacts such as drains or specific body positioning for the X-ray scan.

To better illustrate the most important features of analysed samples in the view of the two considered CNN classifiers a state-of-the-art AI-explainable approach known as Image-Specific Class Saliency \cite{saliency-map-base} is applied.
The maps are obtained for the most accurate model: EfficientNet-B0 trained on the \textit{balanced dataset} (\textit{Results}).
To denoise the results and make them more readable the top 10$\%$ most valuable features were marked on the source image.

\section{Results}

\subsection{GAN-based image generation}

\begin{figure}
    \centering
    \includegraphics[width=0.7\textwidth]{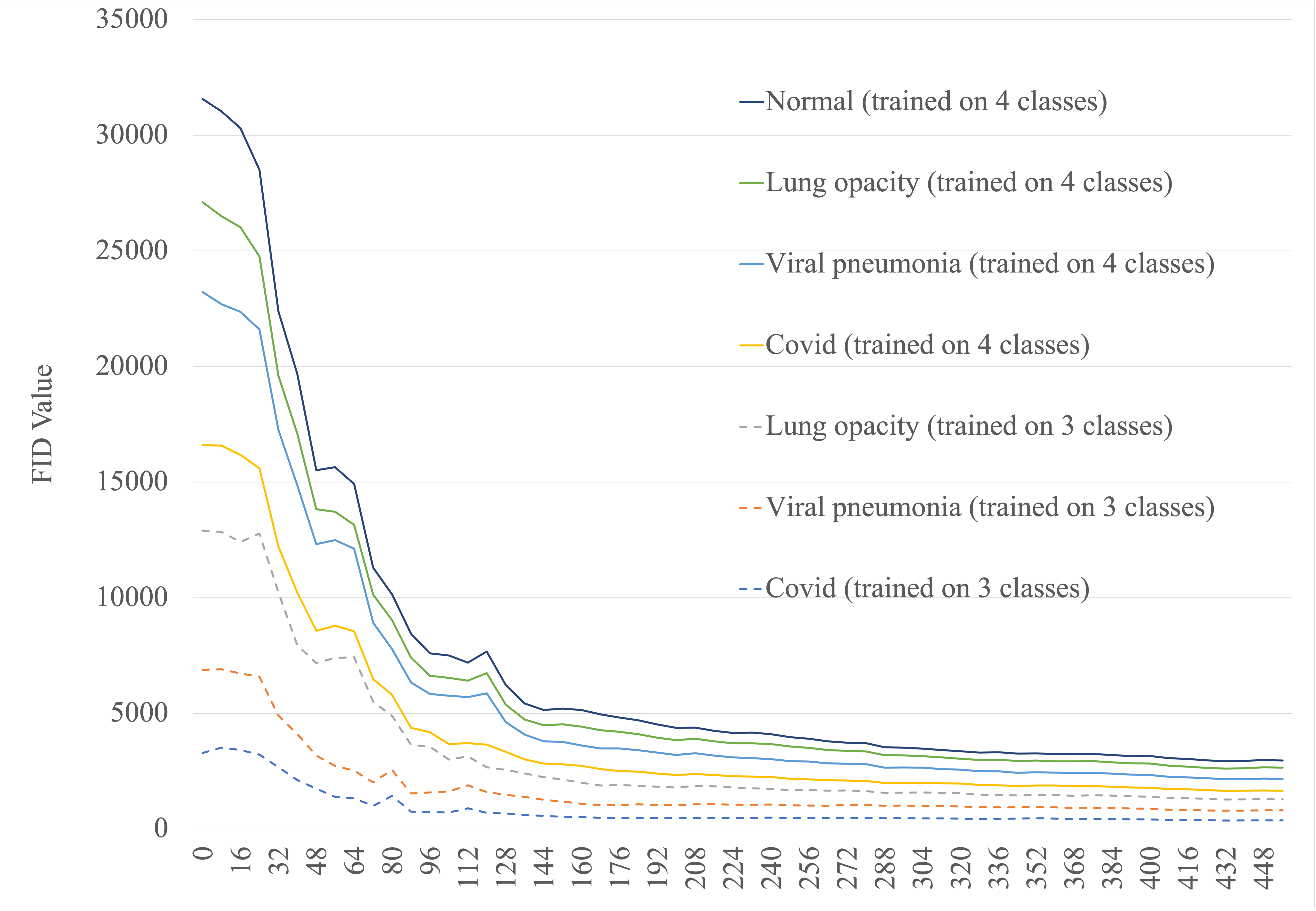}
    \caption{\label{fig:gan_training}FID values as a function of Style-GAN training epoch. Solid lines correspond to the scenario with GAN trained on a sample with four classes (including normal) while dashed lines correspond to a scenario with 3 classes in the training sample. The classes are distinguished with colours according to the legend.}
\end{figure}

\begin{figure}
    \centering
    \includegraphics[width=0.2\textwidth]{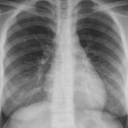}
    \includegraphics[width=0.2\textwidth]{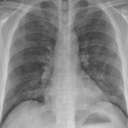}
    \includegraphics[width=0.2\textwidth]{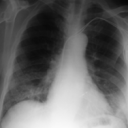}
    \includegraphics[width=0.2\textwidth]{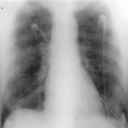}
    \\
    \includegraphics[width=0.2\textwidth]{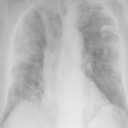}
    \includegraphics[width=0.2\textwidth]{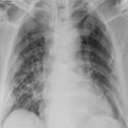}
    \includegraphics[width=0.2\textwidth]{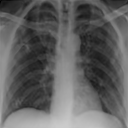}
    \includegraphics[width=0.2\textwidth]{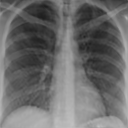}
    \caption{\label{fig:generated_images}Examples of real COVID-19 positive chest X-ray images (top) and examples of GAN-generated COVID-19 chest X-ray images (bottom).}
\end{figure}

The training process of the GAN model is monitored with the FID metric presented in Fig.~\ref{fig:gan_training}.
It is seen that the decrease of FID has reached a plateau for all simultaneously trained classes and the model has learned the characteristics of each of the samples.
It is also seen that including in the training a data set with an additional class (a scenario with four classes including a normal sample), which introduces a large imbalance of the samples, leads to poorer performance of the model.
Final FID values calculated between original and generated samples are presented in Table~\ref{tab:fid_values} together with values calculated between COVID-19 and the other three classes.
It can be observed that the FID values are significantly lower, thus the images are more similar, for generated samples than those obtained between the four classes.
Visual comparison of exemplary generated COVID-19 images to real ones is presented in Fig.~\ref{fig:generated_images}.

\begin{table}
\caption{\label{tab:fid_values}Comparison of FID values calculated between GAN generated samples for three underrepresented samples and corresponding real image sample (first three rows). In the last row, FID values calculated between COVID-19 real image sample and the other real image samples are presented respectively.
}
\begin{ruledtabular}
\begin{tabular}{ccccccccc}
Reference sample    & COVID-19 & Viral pneumonia & Lung opacity & Normal \\
\hline
GAN COVID-19        & $20.9$   &                 &              &        \\
GAN Viral pneumonia &          & $31.8$          &              &        \\
GAN Lung opacity    &          &                 & $38.2$       &        \\
COVID-19            &          & $35.3$          & $173.3$      & $56.5$ \\
\end{tabular}
\end{ruledtabular}
\end{table}

\begin{figure}
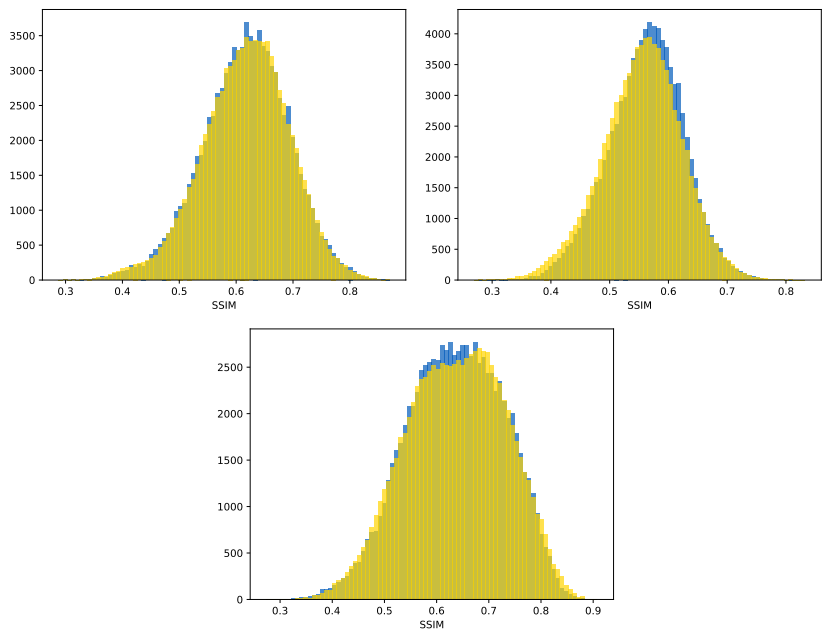

    \centering
    \includegraphics[width=0.33\textwidth,trim={9mm 0 12mm 12mm},clip]{images/comparison_metrics/covid_ssim.pdf}
    \includegraphics[width=0.33\textwidth,trim={9mm 0 12mm 12mm},clip]{images/comparison_metrics/vp_ssim.pdf}
    \includegraphics[width=0.33\textwidth,trim={9mm 0 12mm 12mm},clip]{images/comparison_metrics/lo_ssim.pdf}
    \caption{\label{fig:metrics_generated_ssim}Distribution of the SSIM image similarity metric. For three samples GAN generated images are compared to real ones: COVID-19 (left), viral pneumonia (middle) and lung opacity (right). The blue distribution corresponds to the intra-similarity of images in the real sample and the yellow distribution corresponds to the inter-similarity between generated and real samples.}
\end{figure}

The quality of generated images is also confirmed with classical similarity metrics.
The distributions of those similarity measures between generated and real images are compared for each sample with the distribution obtained within the real samples \ref{fig:metrics_generated}.
The distributions for generated samples describe very well the distributions for corresponding real images.
For an example see the results for the SSIM metric presented in Fig.~\ref{fig:metrics_generated_ssim}.
The small visible differences are almost insignificant compared to differences observed between the classes in Fig.~\ref{fig:metrics_samples}.

\subsection{Classification results}

\begin{figure}
    \centering
    \subfloat[\label{fig:effnet_results_a}]{
        \includegraphics[width=0.45\textwidth]{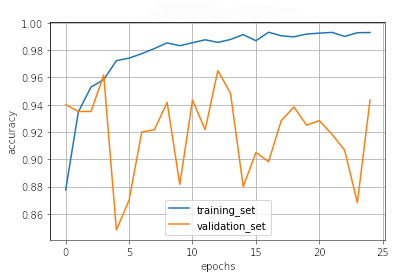}
    }
    \subfloat[\label{fig:effnet_results_b}]{
        \includegraphics[width=0.45\textwidth]{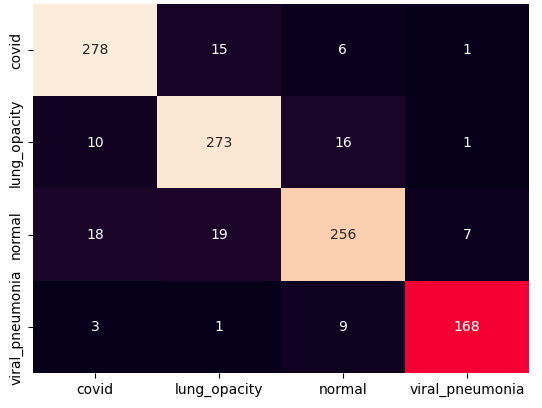}
    }
    \caption{\label{fig:effnet_results}Evolution of training and validation accuracy (left). The test subsample confusion matrix (right) for the best classifier model: EfficientNet-B0 trained on the \textit{balanced dataset}.}
\end{figure}

The training procedure in all scenarios has converged.
Looking at the evolution of training and validation accuracy (Fig.~\ref{fig:effnet_results_a}) indicates that a larger validation dataset would be beneficial.
To stabilise the classifier training a procedure of multiple training runs to select the best model is used (\textit{Materials and methods}).
Resulting classifiers are able to discriminate between classes with good precision, for example, see the confusion matrix of the best overall model presented in Fig.~\ref{fig:effnet_results_b}.
It can also be seen that for the normal class, the rate of misclassified images as COVID-19 ones is higher than for other classes.
This effect is present for all models although smaller for the ones based on the EfficientNet-B0 architecture.
In addition, the EfficientNet-B0 variants excel if the correct classification of viral pneumonia class images with the smallest rate of errors.
However, the EfficientNet-B0 \textit{GAN augmentation} scenario shows increased misclassification of healthy images which leads to a poor performance.
The evolution of the training accuracy as well as confusion matrices for all models are summarised at the end of this section.

The results of the best-trained model for each of the four scenarios are analysed using a number of popular metrics.
As presented in Table~\ref{tab:results} the most accurate model, based on multiple metrics, is an EfficientNet-B0 trained on the \textit{balanced dataset}.
Obtained accuracy of $90.2 \%$ is in line with results obtained by Khan E. et al.~\cite{KhanE_2022} i.e. $92\%$ accuracy using classical augmentation and standard transfer learning with EfficientNet-B1 or $96.13\%$ accuracy with a modified approach.

In the case of Inception-v3, all scenarios that lead to a more balanced dataset have a significant advantage over the raw unbalanced sample.
Including \textit{GAN augmentation} where an accuracy increase of $2\%$ points is observed.
The most accurate classifier is obtained in a \textit{classical augmentation} scenario.
For EfficientNet-B0 result obtained without augmentation is very precise and only a small improvement is observed from the balancing of the training data.
The inclusion of GAN-based augmentation leads to a deterioration of accuracy.

To verify the statistical significance of observed differences a Stuart-Maxwell test is performed.
The p-value of the null hypothesis that EfficientNet-B0 trained on the \textit{balanced dataset} and EfficientNet-B0 trained with \textit{GAN augmentation} return identical answers is equal to $2.12\cdot10^{-09}$ for 3 degrees of freedom.
It is found that p-values for all combinations, besides three cases, are well below $5\%$.
Therefore, differences between metrics observed in Table~\ref{tab:results} are statistically significant.
All p-values are presented in Table~\ref{tab:pvalues} (\textit{Results} section).


\begin{table}
\caption{\label{tab:results}Classification performance achieved for the multi-class problem. Four augmentation scenarios are considered: no~augmentation, balanced dataset, classic augmentation and GAN augmentation. Here f1:~F-Score and mcc:~Matthews Correlation Coefficient. Bold values indicate superior performance.
}
\begin{ruledtabular}
\begin{tabular}{cccccccc}
                &                      & accuracy    & precision   & recall      & f1          & specificity & mcc \\
\hline
Inception-v3    & no augmentation      & $0.855$     & $0.859$     & $0.860$     & $0.860$     & $0.951$     & $0.810$ \\
                & balanced dataset     & $0.883$     & $0.889$     & $0.887$     & $0.888$     & $0.950$     & $0.848$ \\
                & classic augmentation & $0.891$     & $0.896$     & $0.893$     & $0.895$     & $0.963$     & $0.858$ \\
                & GAN augmentation     & $0.871$     & $0.886$     & $0.874$     & $0.880$     & $0.956$     & $0.836$ \\
\hline
EfficientNet-B0 & no augmentation      & $0.895$     &$\bm{0.908}$ & $0.898$     & $0.903$     & $0.964$     & $0.866$ \\
                & balanced dataset     &$\bm{0.902}$ & $0.907$     &$\bm{0.905}$ &$\bm{0.906}$ &$\bm{0.966}$ &$\bm{0.872}$ \\
                & classic augmentation & $0.895$     & $0.904$     & $0.900$     & $0.902$     & $0.964$     & $0.866$ \\
                & GAN augmentation     & $0.841$     & $0.860$     & $0.852$     & $0.856$     & $0.946$     & $0.802$ \\
\end{tabular}
\end{ruledtabular}
\end{table}

\begin{figure}[ht!]
    \centering
    \subfloat[]{
        \includegraphics[width=0.2\textwidth]{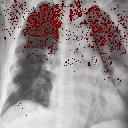}
    }
    \hfill
    \subfloat[]{
        \includegraphics[width=0.2\textwidth]{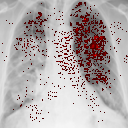}
    }
    \hfill
    \subfloat[]{
        \includegraphics[width=0.2\textwidth]{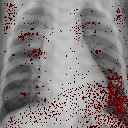}
    }
    \hfill
    \subfloat[\label{fig:saliency_d}]{
        \includegraphics[width=0.2\textwidth]{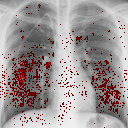}
    }
    \\
    \hfill
    \subfloat[\label{fig:saliency_e}]{
        \includegraphics[width=0.2\textwidth]{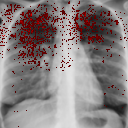}
    }
    \hfill
    \subfloat[\label{fig:saliency_f}]{
        \includegraphics[width=0.2\textwidth]{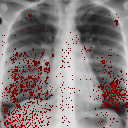}
    }
    \hfill~
    \caption{\label{fig:saliency}Top row - examples of the image-specific saliency maps of the predicted class on the correctly classified images: a) covid, b) lung opacity, c) viral pneumonia, d) healthy lungs. Bottom row - example of saliency maps for incorrectly classified COVID-19 positive image as a healthy one: a) saliency map for COVID-19 class prediction, b) saliency map for healthy class prediction.}
\end{figure}

To validate that results are not biased by models focusing on non-biomarkers saliency maps are used.
The regions of interest of correctly TOP-1 predicted class on the representative test images are shown in Fig.~\ref{fig:saliency}.
It is clear that the saliency maps often indicate the lung region as important, moreover, they do not concentrate on artificial indicators like the drain visible in Fig.~\ref{fig:saliency_d}.
That suggests that the model takes into account genuine lung lesions.

\begin{figure}
    \centering
    \subfloat[]{
        \includegraphics[width=0.38\textwidth]{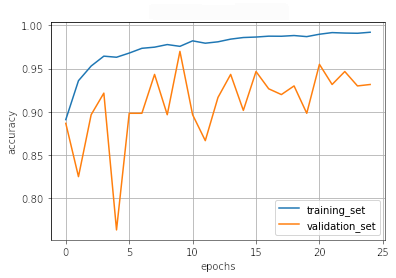}
    }
    \subfloat[]{
        \includegraphics[width=0.38\textwidth]{images/results/efficient_net/accuracy_balanced.png}
    }
    \\
    \subfloat[]{
        \includegraphics[width=0.38\textwidth]{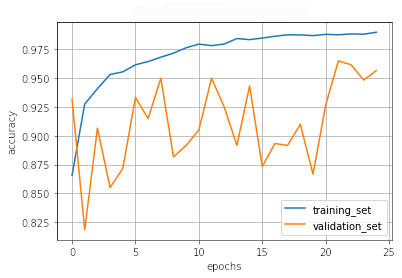}
    }
    \subfloat[]{
        \includegraphics[width=0.38\textwidth]{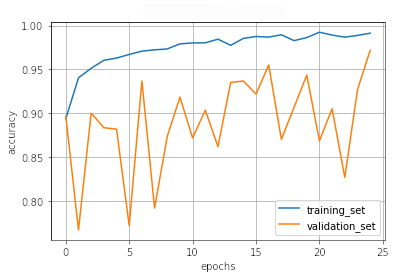}
    }
    \caption{\label{fig:accuracy_training_1}Evolution of training and validation accuracy for Inception-v3. Results are presented for four training scenarios: a) no augmentation, b) balanced dataset, c) classical augmentation, and d) GAN augmentation.}
\end{figure}

\begin{figure}
    \centering
    \subfloat[]{
        \includegraphics[width=0.38\textwidth]{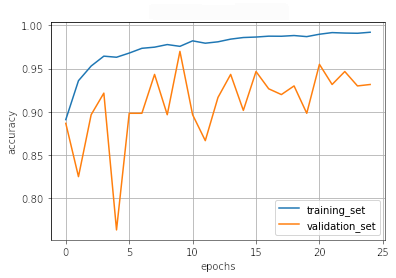}
    }
    \subfloat[]{
        \includegraphics[width=0.38\textwidth]{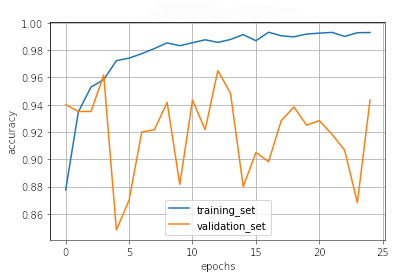}
    }
    \\
    \subfloat[]{
        \includegraphics[width=0.38\textwidth]{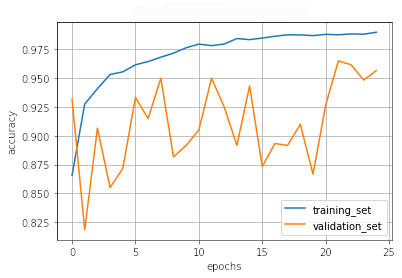}
    }
    \subfloat[]{
        \includegraphics[width=0.38\textwidth]{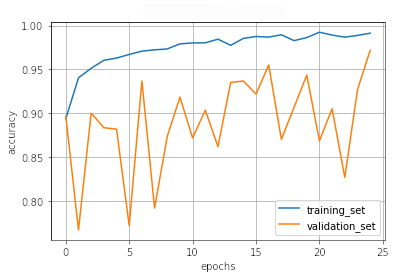}
    }
    \caption{\label{fig:accuracy_training_2}Evolution of training and validation accuracy for EfficientNet-B0. Results are presented for four training scenarios: a) no augmentation, b) balanced dataset, c) classical augmentation, and d) GAN augmentation.}
\end{figure}

\begin{figure}
    \centering
    \subfloat[]{
        \includegraphics[width=0.48\textwidth]{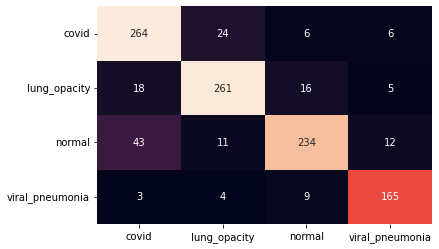}
    }
    \subfloat[]{
        \includegraphics[width=0.48\textwidth]{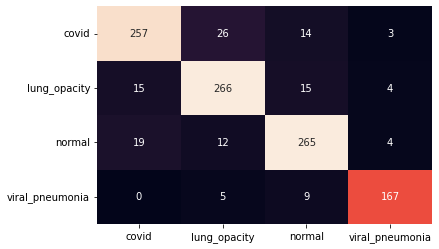}
    }
    \\
    \subfloat[]{
        \includegraphics[width=0.48\textwidth]{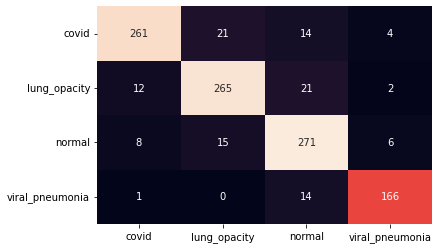}
    }
    \subfloat[]{
        \includegraphics[width=0.48\textwidth]{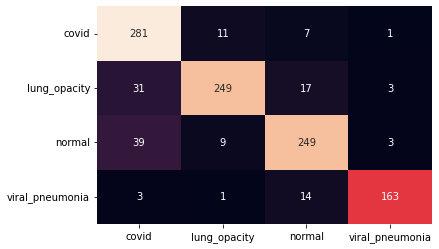}
    }
    \caption{\label{fig:confusion_matrices_inception}Test subsample confusion matrices for Inception-v3. Results are presented for four training scenarios: a) no augmentation, b) balanced dataset, c) classical augmentation, d) and GAN augmentation.}
\end{figure}

\begin{figure}
    \centering
    \subfloat[]{
        \includegraphics[width=0.4\textwidth]{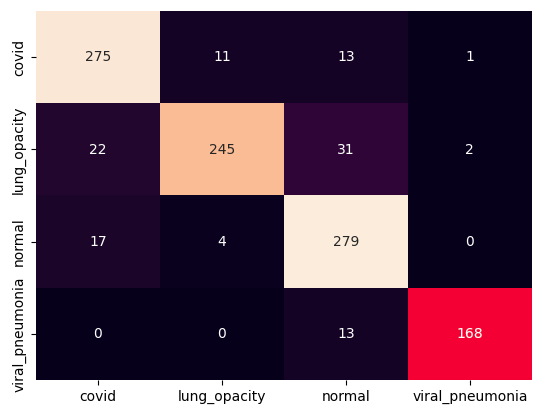}
    }
    \subfloat[]{
        \includegraphics[width=0.4\textwidth]{images/results/efficient_net/balanced_cm_ext.png}
    }
    \\
    \subfloat[]{
        \includegraphics[width=0.4\textwidth]{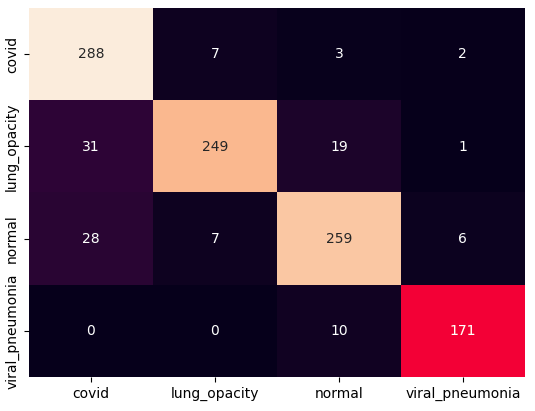}
    }
    \subfloat[]{
        \includegraphics[width=0.4\textwidth]{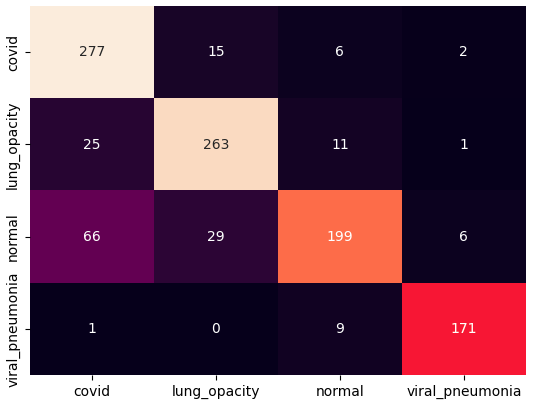}
    }
    \caption{\label{fig:confusion_matrices_effnet}Test subsample confusion matrices for EfficientNet-B0.  Results are presented for four training scenarios: a) no augmentation, b) balanced dataset, c) classical augmentation, d) and GAN augmentation.}
\end{figure}

Training results for the two architectures are summarised in Figs.~\ref{fig:accuracy_training_1}, \ref{fig:accuracy_training_2}, \ref{fig:confusion_matrices_inception}, \ref{fig:confusion_matrices_effnet}, and Table~\ref{tab:pvalues}.
For each architecture four scenarios are considered as described in \textit{Training process} part of the \textit{Materials and methods} Section.

\begin{table}
\caption{\label{tab:pvalues}P-values of pairwise Stuart-Maxwell tests for all four considered scenarios with both Inception-v3 and EfficientNet-B0 models. All tests are performed on the test subsample (\textit{Dataset}) with degrees of freedom equal to 3.
}
\begin{ruledtabular}
\begin{tabular}{ccccccccc}
 & \multicolumn{4}{c}{Inception-v3} & \multicolumn{4}{c}{EfficientNet-B0} \\
 & no aug. & balanced & classic & GAN & no aug. & balanced & classic & GAN \\
\hline
Inception-v3 & &  &  &  &  &  &  &  \\
no aug.      & $1.00$ &  &  &  &  &  &  &  \\
balanced     & $1.50\cdot10^{-04}$ & $1.00$ &  &  &  &  &  &  \\
classic      & $6.64\cdot10^{-08}$ & $2.53\cdot10^{-01}$ & $1.00$ &  &  &  &  &  \\
GAN          & $4.61\cdot10^{-05}$ & $6.10\cdot10^{-09}$ & $1.13\cdot10^{-10}$ & $1.00$ &  &  &  &  \\
\hline
EfficientNet-B0 & &  &  &  &  &  &  &  \\
no aug.         & $5.89\cdot10^{-11}$ & $2.44\cdot10^{-07}$ & $2.97\cdot10^{-05}$ & $8.54\cdot10^{-07}$ & $1.00$ &  &  &  \\
balanced        & $4.36\cdot10^{-02}$ & $1.54\cdot10^{-01}$ & $3.26\cdot10^{-04}$ & $5.83\cdot10^{-06}$ & $1.75\cdot10^{-08}$ & $1.00$ &  &  \\
classic         & $1.03\cdot10^{-04}$ & $5.43\cdot10^{-07}$ & $1.28\cdot10^{-09}$ & $1.30\cdot10^{-01}$ & $4.13\cdot10^{-06}$ & $4.56\cdot10^{-06}$ & $1.00$ &  \\
GAN             & $5.72\cdot10^{-05}$ & $6.17\cdot10^{-14}$ & $0.00$ & $3.22\cdot10^{-08}$ & $0.00$ & $2.12\cdot10^{-09}$ & $2.06\cdot10^{-11}$ & $1.00$ \\
\end{tabular}
\end{ruledtabular}
\end{table}

\section{Conclusion}

In this study, we evaluated the effect of different augmentation techniques on the classification of chest X-ray images.
We consider the multi-class classification problem of chest X-ray images including the COVID-19 positive class, which has been explored only in a handful of published articles.
The proposed StyleGAN2-ADA-based augmentation proved to be beneficial compared to a \textit{no augmentation} scenario when using the Inception-v3 classifier.
For EfficientNet-B0 utilisation of GAN-based augmentation leads to deterioration of performance.
While such comparison is important and common in the literature it is not sufficient to judge GAN applicability for augmentation tasks.
Thus we compared the performance to simple dataset balancing and classical augmentation.
For both classifiers balancing the training dataset is crucial.
While for Inception-v3 \textit{classical augmentation} leads to additional performance gain it is not the case for EfficientNet-B0 where results are comparable to the \textit{no augmentation} scenario.
In both cases, \textit{GAN augmentation} proved to be less performing compared to standard approaches.
The statistical significance of observed differences between considered scenarios was verified with a Stuart-Maxwell test.

The most accurate model based on all considered metrics was an EfficientNet-B0 trained on a balanced dataset with an accuracy of $90.2 \%$.
The EfficientNet-B0 model provided more precise results than Inception-v3 ones for all considered scenarios.
The obtained classification accuracy is in line with the values present in the literature.
The EfficientNet-B0 results indicate that the model is sophisticated enough to extract additional features for the considered dataset size, hence additional augmentation is not beneficial.
To our best knowledge, this is the first result that consistently treats four classes present in COVID-19 chest X-ray datasets with
an approach including GAN-based augmentation.

We took care to verify that synthetic images generated by GAN augmentation are representative of the originals.
The FID metric commonly used in GAN quality assessment was used to monitor the training process and in particular to select the best model.
In addition, we proposed to verify the quality of synthetic images by comparing intra- and iter-similarity distributions of common image similarity measures (RMSE, SRE and SSIM).
The distributions proved to be sensitive in distinguishing different image classes in the dataset while indicating strong similarity between synthetic and corresponding original image samples.
The class-conditional image generation provided by StyleGAN2-ADA allowed us to better utilise the limited training dataset.
However, tests performed with GAN training on a sample heavily skewed by the addition of normal images indicated that proper balancing of GAN input data is important.
Even though an adaptive augmentation is used for StyleGAN2-ADA training the FID metric has significantly improved in the case with only three image classes.
While all the image quality metrics indicate that GAN generated synthetic images are similar to the originals, the performance of \textit{GAN augmentation} is not satisfactory.
This indicates a need to develop additional metrics that would be able to better measure presence of finer details and lack of artefacts in GAN-based images.

We successfully validated that the models did not learn so-called "shortcuts" using saliency maps.
However, the model returns some areas of concern, especially in normal and viral pneumonia classes.
The CNN focuses on the lower parts of the lungs and body parts behind them like a diaphragm, which is definitely not a bio-indicator in this case.
Some of the indicated features can be seen under armpits, a similar result to saliency maps obtained by DeGrave et al.~\cite{sliency-nature-results}.
While such effects could indicate that CNN has learned to recognise "shortcuts", the majority of the attention is focused in the lungs area as one would expect.
To verify this is not a significant effect maps for misclassified images are analysed (Figs.~\ref{fig:saliency_e} and \ref{fig:saliency_e}), e.g image for COVID-19 positive case classified as normal.
No obvious differences from properly classified images are observed.
While the CNNs locate the most important features in the lung area as expected they do not yield straight human interpretable imaging biomarkers.
This highlights the still inseparable black-box limitation of the deep learning algorithms.

Analysis of the results for the EfficientNet-B0-based classifier indicates that the data sample was large enough to obtain satisfactory accuracy without augmentation at all.
Further research will require to verify StyleGAN2-ADA's ability to be trained on even smaller data samples as augmentation is usually more beneficial for small datasets.
An interesting research avenue will be to evaluate how saliency maps correspond to biomarkers present in the images, this will however require manual segmentation of the images by a physician.

\begin{acknowledgments}
This work was completed in part with resources provided by the Świerk Computing Centre at the National Centre for Nuclear Research.
This work benefited from the software tools developed in the frame of the EuroHPC PL Project, Smart Growth Operational Programme 4.2.
We gratefully acknowledge Polish high-performance computing infrastructure PLGrid (HPC Centers: ACK Cyfronet AGH) for providing computer facilities and support within computational grant no. PLG/2022/015617.
\end{acknowledgments}

\nocite{*}
\bibliography{main}

\end{document}